\documentclass[12pt]{article}

\newtheorem{theorem}{Theorem}%  meant for continuous numbers
\newtheorem{corollary}{Corollary}%

\usepackage[utf8]{inputenc}
\usepackage{setspace}
\usepackage[margin=1in]{geometry}
\usepackage{bm}
\usepackage{booktabs}
\usepackage{graphicx,psfrag,epsf}
\usepackage{multirow}
\usepackage{threeparttable}
\usepackage{algpseudocode}
\usepackage{algorithm}
\usepackage{url}
\usepackage{amsmath}
\usepackage{natbib}
\usepackage{lineno}
\DeclareMathOperator*{\argmin}{arg\,min}
\algnewcommand{\Input}{\textbf{Input}}
\algnewcommand{\Output}{\textbf{Output}}

\newcommand{\bmf}{\bm{f}}
\newcommand{\bmg}{\bm{g}}

\newcommand{\bmA}{\bm{A}}
\newcommand{\bmB}{\bm{B}}
\newcommand{\bmC}{\bm{C}}
\newcommand{\bmD}{\bm{D}}
\newcommand{\bmG}{\bm{G}}
\newcommand{\bmH}{\bm{H}}
\newcommand{\bmX}{\bm{X}}
\newcommand{\bmS}{\bm{S}}
\newcommand{\bmP}{\bm{P}}

\newcommand{\bmR}{\bm{R}}

\newcommand{\bmU}{\bm{U}}

\newcommand{\bmW}{\bm{W}}

\newcommand{\bmF}{\bm{F}}

\newcommand{\bmlambda}{\bm{\lambda}}
\newcommand{\bmbeta}{\bm{\beta}}
\newcommand{\bmtheta}{\bm{\theta}}

\newcommand{\bmepsilon}{\bm{\epsilon}}

\newcommand{\bmzero}{\bm{0}}

\newcommand{\bmGamma}{\bm{\Gamma}}
\newcommand{\bmSigma}{\bm{\Sigma}}
\newcommand{\bmLambda}{\bm{\Lambda}}
\newcommand{\bmPhi}{\bm{\Phi}}

\newcommand{\bmTheta}{\bm{\Theta}}

\newcommand{\tildeY}{\tilde{Y}}

\newcommand{\tildebmD}{\tilde{\bm{D}}}

\newcommand{\tildebmX}{\tilde{\bm{X}}}

\newcommand{\hatbmbeta}{\hat{\bm{\beta}}}
\newcommand{\hatbmtheta}{\hat{\bm{\theta}}}
\newcommand{\hatbmlambda}{\hat{\bm{\lambda}}}

\newcommand{\hatbmPhi}{\hat{\bm{\Phi}}}
\newcommand{\hatbmTheta}{\hat{\bm{\Theta}}}

\newcommand{\hatp}{\hat{p}}
\newcommand{\hatq}{\hat{q}}

\newcommand{\hatK}{\hat{K}}
\newcommand{\hattheta}{\hat{\theta}}
\newcommand{\hattau}{\hat{\tau}}

\newcommand{\hatbmR}{\hat{\bm{R}}}

\newcommand{\hatbmU}{\hat{\bm{U}}}

\title{A New Integrative Learning Framework for Integrating Multiple Secondary Outcomes into Primary Outcome Analysis: A Case Study on Liver Health}  
\author{Daxuan Deng$^{1}$, Peisong Han$^{2}$, Shuo Chen$^{3}$, Ming Wang$^{4}$, Chixiang Chen$^{3*}$\\
	$^{1}$ Department of Public Health Sciences,\\
    Penn State College of Medicine, Hershey, PA, USA\\ 
    $^{2}$ Biostatistics Innovation Group, Gilead Sciences, Foster City, CA, USA\\
    $^{3}$ Department of Epidemiology and Public Health,\\
    University of Maryland School of Medicine, Baltimore, MD, USA\\
    $^{4}$ Department of Population and Quantitative Health Sciences,\\
    Case Western Reserve University, Cleveland, OH, USA}

\date{}

\providecommand{\keywords}[1]
{
  \small	
  \textbf{\textit{Keywords: }} #1
}

\begin{document}
\maketitle
%\subtitle{Subject Section}

\abstract{In the era of big data, secondary outcomes have become increasingly important alongside primary outcomes. These secondary outcomes, which can be derived from traditional endpoints in clinical trials, compound measures, or risk prediction scores, hold the potential to enhance the analysis of primary outcomes. Our method is motivated by the challenge of utilizing multiple secondary outcomes, such as blood biochemistry markers and urine assays, to improve the analysis of the primary outcome related to liver health. Current integration methods often fall short, as they impose strong model assumptions or require prior knowledge to construct over-identified working functions. This paper addresses these statistical challenges and potentially opens a new avenue in data integration by introducing a novel integrative learning framework that is applicable in a general setting. The proposed framework allows for the robust, data-driven integration of information from multiple secondary outcomes, promotes the development of efficient learning algorithms, and ensures optimal use of available data. Extensive simulation studies demonstrate that the proposed method significantly reduces variance in primary outcome analysis, outperforming existing integration approaches. Additionally, applying this method to UK Biobank (UKB) reveals that cigarette smoking is associated with increased fatty liver measures, with these effects being particularly pronounced in the older adult cohort.
%See below.
}


% \boxedtext{
% \begin{itemize}
% \item Key boxed text here.
% \item Key boxed text here.
% \item Key boxed text here.
% \end{itemize}}

\maketitle

\setstretch{1.5}

%\section*{Abstract}

\section{Introduction}\label{Introduction}
In traditional biomedical research, the primary outcome, also referred to as the primary endpoint in clinical studies, is the most important variable measured to assess whether a treatment or exposure being tested has had a significant effect. Secondary outcomes, on the other hand, are additional variables measured in the same study that provide overlapping or supplementary information about the treatment or exposure, including safety and tolerability, qualify of life, relevant biomarkers, and broader impacts on health and well-being \citep{levit2024critical}. Compared to the primary outcome, secondary outcomes have historically received less attention but have become increasingly seen in contemporary biomedical studies. Here, we summarize three sources where secondary outcomes can be defined:

Type 1 – Clinical endpoints. For example, in a clinical trial or epidemiological study evaluating a new medication for hypertension \citep{egan2016systolic}, the primary outcome could be the reduction in systolic blood pressure after 12 weeks of treatment. Secondary outcomes could include changes in diastolic blood pressure, changes in cholesterol levels, adherence to medication, etc. This is the most classic source of secondary outcomes in the literature. The following two types are less conventional but are increasingly observed in data science research using real-world or omics data.

Type 2 – Compound indices defined by experts or leveraging over multi-domain diseases. For example, in studies of older adults, individuals may be classified as frail if they meet three or more of the following criteria: unintentional weight loss, exhaustion, weakness, slow walking speed, and low physical activity \citep{bandeen2015frailty}. As a result, in addition to frailty as the primary outcome to study the vulnerability of older adults, the presence of each individual component can be regarded as a secondary outcome. 
% \red{I think type 2 and type 3 are similar here, where the former weight is fixed the latter one is estimated from data-driven approaches. I think you can keep Type 3 because this is relevant to data application. You can also think about the multi-domain disease (neuro-degenerative disorders with both motor and non-motor systems)}

Type 3 – Data-driven risk scores. For example, the neurobiological risk score is a composite measure that quantifies an individual's risk of developing a neurological or psychiatric disorder based on a combination of T1-weighted and diffusion tensor imaging measures \citep{rodrigue2023multimodal}. This score can be constructed by weighting each factor based on its relative importance and contribution to the overall risk, with weights estimated using data science tools, including XGBoost \citep{ke2017lightgbm}. Hence, all individual factors used to construct risk scores (the primary outcome) can serve as secondary outcomes. 

Despite being termed ``secondary", these outcomes have the potential to enhance estimation and inference in primary outcome analysis. This is particularly true for the latter two sources, where the primary outcome is a summary score derived from secondary outcomes and thus lacks the granular information captured by each secondary outcome. Relying solely on the primary outcome may not fully exploit the information of biological disease mechanisms or specific phenotypes from individual secondary outcomes. Therefore, there is a strong need to study methods that integrate information from secondary outcomes into the primary outcome analysis.

The case study in this paper focuses on two primary outcomes: liver proton density fat fraction (PDFF), a quantitative imaging biomarker that measures hepatic fat content \citep{xia2023association}, and a new liver-biological risk score (LBRS), derived by modeling liver PDFF using predictive factors, such as blood measurements and urine assays. The latter outcome serves as a surrogate marker for liver PDFF and is particularly useful in scenarios where high-quality image markers require substantial effort and may provide an ``easy pass" for the timely identification of high-risk subjects and assessment of lifestyle behaviors influencing liver health (see Section \ref{Liver-biological risk scores} for more discussion). By studying liver PDFF/LBRS as primary outcomes, respectively, we aim to study the impact of cigarette smoking and explore whether this effect varies across different age groups. While recent studies have linked smoking to chronic liver disease \citep{marti2022cigarette}, its role in fatty liver remains poorly understood. Given the serious health consequences associated with fatty liver disease \citep{xia2023association}, there is a pressing need to investigate this relationship and provide evidence-based guidance for prevention. Beyond clinical relevance, we also pose a methodological question: can secondary outcomes, such as blood biochemistry factors and urine assays, contribute meaningfully to the analysis of the primary outcome? A plausible answer is yes, considering that these secondary outcomes are naturally associated with the primary outcome, even after adjusting for other covariates. However, a rigorous statistical tool to harness this association and robustly integrate information from multiple secondary outcomes remain under-explored.

From a statistical methodology perspective, the main goal of integrating information from secondary outcomes is to improve estimation efficiency and increase statistical power, without compromising estimation consistency, in the analysis of the primary outcome. Achieving this goal poses a significant challenge: how to leverage the association between primary and secondary outcomes without relying on strong assumptions, such as assuming shared effect sizes across outcomes \citep{chen2023efficient}. One possible solution is to specify a joint likelihood for both primary and secondary outcomes by jointly modeling their means and associations \citep{ ezzalfani2019joint, alsefri2020bayesian}. However, this approach is often computationally intensive, especially when outcomes follow distinct distributions. More importantly, it does not differentiate the roles of primary and secondary outcomes, which may yield bias in the primary analysis if the joint model is mis-specified. Therefore, it is desirable to develop methods that preserve robust and unbiased inference of the primary outcome while integrating information from secondary outcomes to improve estimation efficiency in the primary analysis, without treating all outcomes as equally important. In this context, \cite{chen2022improving} proposed a re-weighting estimation scheme called the Vertical Information Scores (VIS) estimator to borrow information from a secondary outcome in a robust and computationally efficient manner. Here, ``robust" means that the mis-specification of the secondary outcome model will not introduce estimation bias to the primary outcome analysis. Further, \cite{chen2023efficient} extended the work to accommodate multiple secondary outcomes, with other variants addressing missing data \citep{deng2024robust} and analyzing survival outcomes \citep{chen2023synthesizing}. Despite these strengths, these methods require a user-specified, over-identified estimating function, where the number of estimating function is larger than the number of parameter \citep{owen2001empirical}. 
% while it is common in practice to construct a just-identified estimating function (i.e., equal dimension).
 While over-identification could be achieved in repeatedly measures settings by stacking multiple estimating functions with varying base matrices  \citep{chen2022improving}, it remains unclear for a general data setting, limiting broader applicability. Before describing our solutions, we note that other advanced methods have been developed for integrating external information into internal analysis \citep{chatterjee2016constrained,zhang2020generalized, duan2022heterogeneity,han2024improving, zhai2024integrating, chen2024integrating}. However, these methods are designed for combining independent studies and are not applicable to our setting, where both primary and secondary outcomes are drawn from the same study.

\begin{figure}[ht]
    \centering
    \includegraphics[scale=0.5]{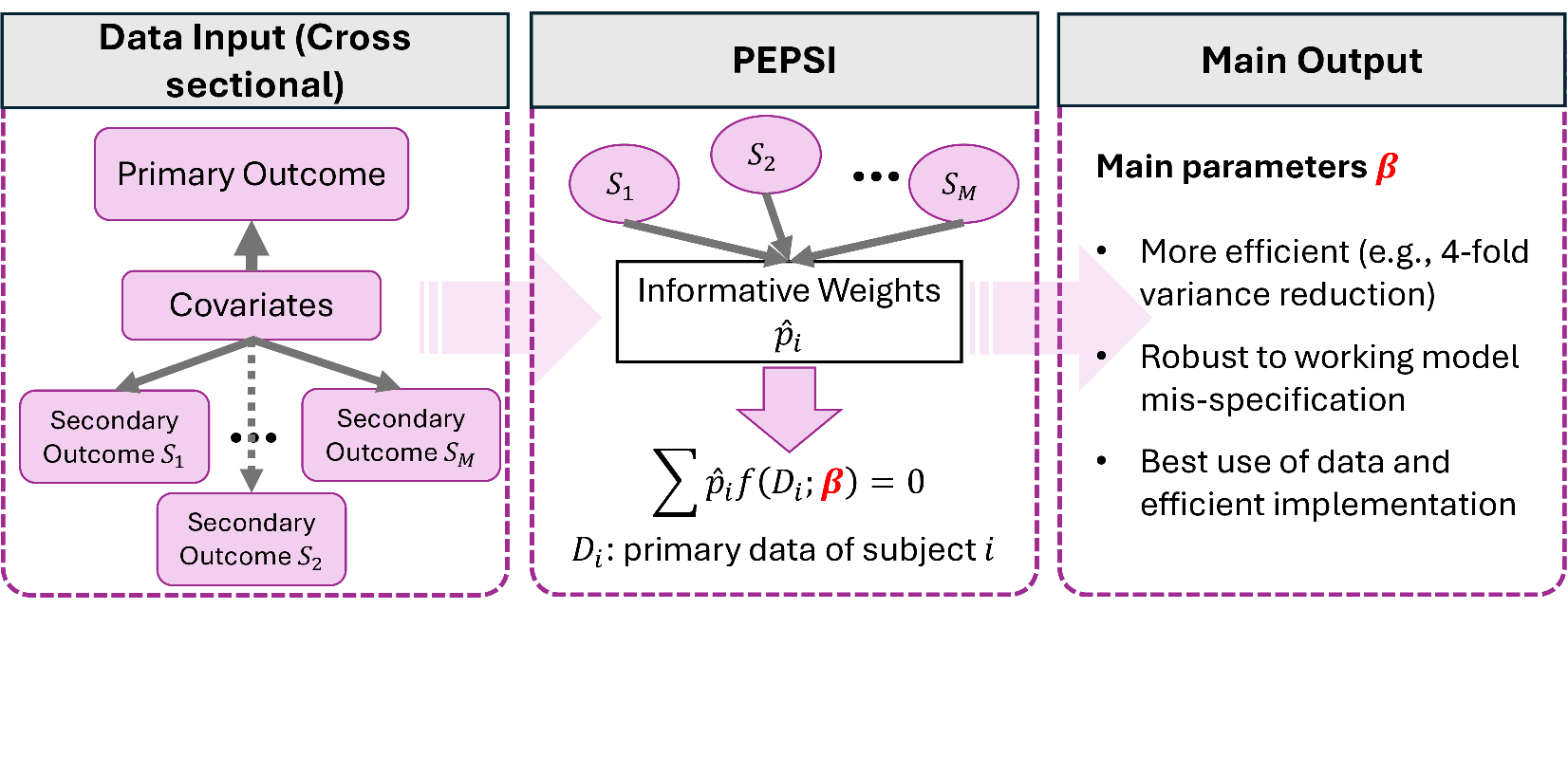}
    \caption{Overall workflow of PEPSI. Dashed line in the left panel implies that covariates have partial or no effect on the secondary outcome.}
    \label{workflow}
\end{figure}

In this paper, we propose a new integrative learning framework named PEPSI, \textbf{P}enalized \textbf{E}mpirical likelihood and \textbf{P}rincipal component analysis-based \textbf{S}econdary outcome \textbf{I}ntegrative learning. PEPSI is designed primarily (but not limited to) for settings with multiple cross-sectional secondary outcomes, and it innovatively integrates data science techniques, including constrained optimization, penalized empirical likelihood, and principal component analysis, to achieve three goals (Figure \ref{workflow}). 1) PEPSI robustly integrates information from secondary outcomes in a data-driven manner, without the need for prior knowledge to specify over-identified functions, a major limitation of existing methods \citep{wolf2024commentary, chen2023efficient}. The key to achieving this is learning zero coefficients in secondary outcome models. Notably, efficiency gains occur when some coefficients in certain secondary models are zero, but users do not need to know which coefficients from which secondary outcome models are zeros. Even when all covariate effects are non-zero, PEPSI performs at least as well as methods without data integration. As a result, PEPSI preserves the original estimand or the hypothesis of interest defined by the primary outcome analysis or study protocol. 2) PEPSI efficiently accommodates multiple secondary outcomes through a computationally efficient learning algorithm. 3) Within certain classes of estimators, PEPSI optimally integrate multiple secondary outcomes, demonstrating superior performance over existing integration schemes \citep{chen2023efficient}. Substantial simulation studies also demonstrate PEPSI's advantages, showing substantial reductions in estimation variability with negligible estimation bias. Moreover, the method significantly enhances the analysis studying the impact of cigarette smoking on liver health. Notably, compared with the classic estimator, PEPSI achieves over $4$-fold relative efficiency in many estimates and identifies several significant factors influencing the LBRS outcome, such as age $\geq 65$ and its interaction with smoking, suggesting heterogeneous smoking effects. 
% The following manuscript is organized as follows: Section \ref{Preliminaries} provides the scientific context of our case study and the detailed setting for PEPSI; Section \ref{PEPSI} details the PEPSI framework; Section \ref{simulation} presents numerical evaluations of the proposed method; Section \ref{Application} offers data analysis and results interpretation; and Section \ref{Discussion} discusses the data application, method extensions, and concludes the paper.
% As described in Section \ref{Introduction}, biological risk scores have recently gained popularity for predicting disease onset and studying disease mechanisms by examining the risk factors driving unfavorable scores. Typically, this index is formulated by regressing phenotype or image markers on biological assays using machine learning models, serving as an informative prediction score for subjects where rigorous disease diagnosis or high-quality image markers require substantial effort. 
\section{Preliminaries}\label{Preliminaries}
\subsection{Liver-Biological Risk Scores} \label{Liver-biological risk scores}
Recent literature revealed that PDFF was linked to  significantly elevated risks for liver cancer, non-alcoholic steatohepatitis, cirrhosis, and other forms of liver disease \citep{xia2023association}. However, obtaining accurate liver PDFF, captured by MRI-based water-fat separation technique \citep{starekova2021quantification}, can be time-consuming, expensive, and is known to have high sampling variability \citep{kramer2017accuracy}. On the other hand, blood biochemistry measures and urine biomarkers are much easier and cheaper to obtain and have been shown to be associated with chronic liver conditions \citep{longo2012harrisons,kwo2017acg}. Therefore, as one of the application goals in this paper, we are interested in constructing LBRS and studying their utility compared to the liver PDFF measure. As illustrated in Section \ref{Application}, we constructed the LBRS by predicting PDFF from multiple blood and urine biomarkers using a machine learning model. The constructed LBRS can also be interpreted as a de-noised outcome by projecting liver PDFF onto the space of informative biomarkers. This risk score may capture the underlying information of liver PDFF explained by the biomarkers and thus has the potential to provide an ``easy pass" and possibly strong signals for the timely identification of high-risk subjects, assessment of environmental factors and lifestyle behaviors influencing LBRS, and examination of their heterogeneous effects among subcohorts without requiring the availability of liver PDFF measures. The results will also inform future trial designs to infer definitive conclusions.

After constructing LBRS, our clinical interest is to study and compare the impact of cigarette smoking on LBRS and liver PDFF, and to explore whether this effect differs across age groups, a topic that remains understudied in the literature. Therefore, our primary outcomes are LBRS and liver PDFF, with cigarette smoking as the primary exposure. Our secondary outcomes include $22$ blood biochemistry measures and $3$ urine assays, which were selected by domain experts and are expected to be strongly associated with the two primary outcomes. Given the presence of multiple secondary outcomes, our methodological objective is to develop an effective integrative learning strategy that enhances estimation efficiency in the primary analysis by robustly leveraging information from these secondary outcomes.

\subsection{Notations and Basic Set-up}\label{Notations and Basic Set-up}
We now formally introduce the primary and secondary data in a generic setting, of which our data application is one special scenario. Specifically, let $\bmD_{i} = \{Y_i, \bmX_i\}$, $i=1,2,\dots,n$, be independent and identically distributed (i.i.d.) primary data for the primary analysis, where $Y$ and $\bmX$ are the primary outcome and covariate vector, respectively, and $n$ is the sample size. We denote $\hat\bmbeta$ as the estimated parameter vector, which can be obtained by solving the estimating equation $\sum_{i=1}^n \bmf(\bmD_{i}; \bmbeta) = \bmzero$. The function $\bmf(\cdot)$ can be any regular estimating function, such as the first derivative of a least squares loss function, a score function, or a function from generalized estimating equations \citep{van2000asymptotic}. The limiting value of $\hat\bmbeta$, denoted by $\bmbeta_0$, satisfies $E\{\bmf(\bmD; \bmbeta_0)\} = \bmzero$. In Section \ref{Liver-biological risk scores}, $\bmbeta$ contains the effects of risk factors, such as cigarette smoking, age, and their interaction term, on LBRS/PDFF.

On the other hand, we may have $M \geq 1$ i.i.d. secondary data $\tildebmD_{mi} = \{\tildeY_{mi}, \tildebmX_{mi}\}$, for $i=1,2,\dots,n$ and $m=1,\ldots,M$, collected from the same cohort of participants, where $\tildeY_m$ and $\tildebmX_m$ denote the secondary outcome and covariate vector in the $m$-th secondary data, respectively. We emphasize that the primary and secondary data are not independent of each other. In our case study, for instance, $\tildeY_m$ includes blood and urine biomarkers that are believed to be associated with and predictive of liver health (e.g., LBRS/PDFF). Throughout this paper we focus on the setting where both primary and secondary data are cross-sectional and share the same covariates, i.e., $\bmX = \tildebmX_m$, for all $m=1,\ldots,M$, which aligns with our application. However, the proposed method can be extended to more general settings.

Before introducing the method, we briefly explain how the variables described above are typically determined in practice. 
% The primary and secondary analysis are defined sequentially and are mutually exclusive.
The primary outcome and the corresponding hypothesis should be defined prior to the analysis stage, while secondary outcomes are typically pre-specified by the study protocol or defined from the construction of the primary outcome. In contrast, covariates should be selected independent of secondary outcomes, based on the original research objectives related to the primary outcome. In our case study, the primary outcome was constructed using a set of informative biomarkers selected by clinical collaborators. Consequently, the secondary outcomes naturally consisted of these variables. This selection process took place during the design phase and was informed by domain expertise. We refer readers to Section \ref{Multiple Secondary Data} for general guidelines on selecting secondary outcomes. Covariates, on the other hand, were chosen according to the specific research question. For instance, if examining sex differences in liver PDFF is of interest, sex would be included as a covariate and excluded from the pool of secondary outcomes.

In the following section, we describe a novel integrative learning approach that incorporates information from all secondary data into the primary data analysis in a data-driven and robust manner, without altering the original estimand or the hypothesis related to $\bmbeta_0$ defined by the primary outcome analysis
or study protocol. To smoothly introduce our method, we start with the setting where there is only one secondary data and then illustrate the setting of multiple secondary data.

\section{The PEPSI Framework}\label{PEPSI}

% Suppose we have $M$ secondary data $\{\tildebmD_m\}_{m=1}^M$. For illustration, we start with integrating single secondary data in Section \ref{Single Secondary Data}, then we generalize it to multiple secondary data in Section \ref{Multiple Secondary Data}.

\subsection{Single Secondary Data}\label{Single Secondary Data}
We first outline the learning procedure and then explain the underlying rationale behind the method. Suppose there is only one secondary data available in the study, denoted by $\tildebmD_m$, for some $m$ satisfying $1 \le m \le M$. To enhance parameter estimation and inference by integrating information from the secondary data, we propose a new estimator, named the Penalized empirical likelihood-based Secondary outcome Integrative learning (PSI) estimator, a special case of PEPSI, denoted by $\hatbmbeta_{PSI}$, which is obtained by solving the following re-weighted estimating equation:
\begin{equation} \label{weighted estimating equations}
\sum_{i=1}^{n}\hatp_{mi}\bmf(\bmD_{i};\bmbeta)=\bmzero.
\end{equation}
The weights are proposed to be $\hatp_{mi} = n^{-1}\{1 + \bmlambda_m^T(\hatbmtheta_m)\bmg_m(\tildebmD_{mi}; \hatbmtheta_m)\}^{-1}$, for $i = 1, \ldots, n$, where $\bmlambda_m(\bmtheta_m)$ is a function of $\bmtheta_m$ satisfying $n^{-1}\sum_{i=1}^n\{1+\bmlambda_m^T\bmg_m(\tildebmD_{mi};\bmtheta_m)\}^{-1} \linebreak \bmg_m(\tildebmD_{mi};\bmtheta_m)=\bmzero$. The function $\bmg_m(\tildebmD_{m}; \bmtheta_m)$ is a ``just/over-identified" working estimating function for the secondary data, parameterized by $\bmtheta_m$, which satisfies $E\{\bmg_m(\tildebmD_{m};\linebreak \bmtheta_m^\ast)\} = \bmzero$ for some value $\bmtheta_m^\ast$. Let the dimensions of $\bmg_m(\tildebmD_{m}; \bmtheta_m)$ and $\bmtheta_m$ be $r_m$ and $p_m$, respectively. Here, ``just/over-identified" means that the length of the estimating function vector is equal to or greater than the length of the parameter vector, i.e., $r_m\ge p_m$ \citep{owen2001empirical}. In practice, an over-identified estimating function within a regression framework can be constructed by setting certain components of $\hat{\bm{\theta}}_m$ to exactly zero when it is known that the corresponding covariates have no effect on the secondary outcome. However, we highlight here that prior knowledge of constructing the over-identified function is not required to implement the PSI framework.

Moreover, $\hatbmtheta_m$ is obtained by minimizing the following penalized loss function \citep{leng2012penalized}:
\begin{equation}\label{penalized loss}
    l_{pm}(\bmtheta_m)=\sum_{i=1}^n\log\{1+\bmlambda_{m}^T(\bmtheta_m)\bmg_m(\tildebmD_{mi};\bmtheta_m)\} +n\sum_{j=1}^{p_m}\rho_{\tau_m}(|\theta_{mj}|).
\end{equation}
Here, the first term $l_m(\bmtheta_m)=\sum_{i=1}^n\log\{1+\bmlambda_{m}^T\bmg_m(\tildebmD_{mi};\bmtheta_m)\}$ corresponds to the negative logarithm of the empirical likelihood ratio function \citep{qin1994empirical}, while the second term involves a penalty function $\rho_{\tau_m}(\cdot)$ with a tuning parameter $\tau_m$, where $\{\theta_{mj}\}_{j=1}^{p_m}$ are the entries of the vector $\bmtheta_m$. We sometimes write $\hatbmtheta_m=\hatbmtheta_{m,\tau_m}$ and $l_{pm}(\bmtheta_m)=l_{pm}(\bmtheta_{m,\tau_m})$ to emphasize the dependency on $\tau_m$. In this paper, we use the SCAD penalty \citep{fan2001variable} for illustration, though other penalty functions can be applied \citep{zou2006adaptive,zhang2010nearly}. Therefore, the proposed weights $\hatp_{mi}$ can be interpreted as an extension of the empirical likelihood framework \citep{qin1994empirical}, where the parameters are estimated using a penalized procedure.

We now provide an intuitive rationale for how the proposed PSI estimator works: The constructed weights $\{\hatp_{mi}\}_{i=1}^n$ summarize useful information from the secondary data and serve as a bridge connecting the primary and secondary data, while the proposed penalized optimization aims to identify zero-valued parameters in $\bmtheta_m$. When the function $\bmg_m(\tildebmD_{mi};\bmtheta_m)$ is originally just identified and the penalized regression does not identify any zero-valued parameters in $\bmtheta_m$, the weights reduce to $1/n$, i.e., non-informative weights \citep{qin1994empirical,chen2022improving}. Under this scenario, there is no efficiency gain for estimating $\bmbeta$. However, if any zero components of $\bmtheta_m$ are identified, the efficiency gain can be substantial. Intuitively,  shrinkage of very small estimated parameter values to zero implicitly facilitates the construction of over-identified working functions in $\bmg_m(\tildebmD_{mi}; \bmtheta_m)$, which in turn promotes a more efficient estimate of the empirical distribution \citep{qin1994empirical, chen2022improving}. As a result, the weights in \eqref{weighted estimating equations} incorporate additional information from the secondary data, thereby enhancing the estimation of $\bmbeta$ in the primary analysis. It is worth noting that the existence of null effects from certain covariates on the secondary outcome is common in real practice. 
% For example, only a few genetic factors among the whole genome sequence are associated with Alzheimer's disease onset \citep{andrews2023complex} \red{I do not think this example is helpful. You can refer an example related to your motivating data example}. 
In Section \ref{Application}, for example, we observed that smoking had little effect on several secondary outcomes, such as Calcium in blood, Potassium in urine. 
% Cholesterol, C-reactive protein, Gamma glutamyltransferase, Glucose, HDL cholesterol, Phosphate, SHBG, Total bilirubin, Total protein, Triglycerides, Creatinine (enzymatic) in urine, , Sodium in urine.Alkaline phosphatase, Alanine aminotransferase, 
Consistently, existing literature provide limited evidence supporting significant effects of smoking on these outcomes after adjusting for other covariates \citep{breitling2015smoking}. Importantly, regardless of covariate effects in secondary outcome models, our method does not impose any assumptions on the true effect sizes from the primary outcome model, thus maintaining substantial flexibility in application. In summary, the proposed PSI facilitates data integration by auto-constructing over-identified working functions through the data-driven identification of zero components in $\bmtheta_m$. For further details, see Theorem \ref{asymptotic of psi} and the accompanying discussion.

The calculation of the weights $\{\hatp_{mi}\}_{i=1}^n$ involves constrained optimization combined with a penalty, which could be computationally intensive. To promote an efficient computation, we adopt a quadratic approximation for the penalty function in \eqref{penalized loss} and iteratively update $\bmtheta_m$ and $\bmlambda_m$ iteratively. During optimization, we set entries of $\bmtheta_m$ to zero if their magnitudes fall below a pre-specified threshold $\gamma$, which is set to $1\times 10^{-3}$ in both simulations and application. The tuning parameter $\tau_m$ is selected by minimizing the following Bayesian information criteria:
$\text{BIC} (\tau_m)=2l_m(\hatbmtheta_{m,\tau_m}) + C_m\log(n)\sum_{j=1}^{p_m}I(\hattheta_{mj,\tau_m} \ne 0)$,
where $\hatbmtheta_{m,\tau_m}$ are obtained by minimizing \eqref{penalized loss} under $\tau_m$, and $C_m$ is a hyper-parameter controlling the bias-variance trade-off in practice. As suggested by \citet{leng2012penalized}, setting $C_m=\max(\log\log p_m,1)$ provides a reasonable choice. The pseudo algorithm for implementing PSI is summarized in Algorithms \ref{algorithm: PEL} and \ref{algorithm: PSI}. 

\begin{algorithm}[ht]
%\setstretch{1.5}
\caption{Algorithm for obtaining the penalized empirical likelihood estimator}\label{algorithm: PEL}
\begin{algorithmic}
\State \text{Input: secondary data} $\{\tildebmD_{mi}\}_{i=1}^n.$ 
\ForAll {$\tau_m$} 
    \State $\hatbmlambda_m \leftarrow \bmzero.$
    \Repeat 
        \State $\hatbmtheta_{m,\tau_m} \leftarrow \argmin_{\bmtheta_{m,\tau_m}} l_{pm}(\bmtheta_{m,\tau_m}).$ 
        \State \text{Update $\hatbmlambda_m$ by solving $\sum_{i=1}^n\{1+\hatbmlambda_m^T\bmg_m(\tildebmD_{mi};\hatbmtheta_{m,\tau_m})\}^{-1} \bmg_m(\tildebmD_{mi};\hatbmtheta_{m,\tau_m})=\bmzero.$} 
        \ForAll{$j$}
            \State $\hattheta_{mj,\tau_m} \leftarrow \hattheta_{mj,\tau_m} I(|\hattheta_{mj,\tau_m}|\le \gamma).$
        \EndFor
    \Until{$\hatbmtheta_{m,\tau_m}$ \text{converges.}}
\EndFor
\State $\hattau_m\leftarrow \argmin_{\tau_m} \text{BIC} (\tau_m).$
\State $\hatbmtheta_{m} \leftarrow \hatbmtheta_{m,\hattau_m}.$
\State $\text{Output: } \hatbmtheta_{m}.$
\end{algorithmic}
\end{algorithm}

\begin{algorithm}[ht]
\caption{Algorithm for obtaining the PSI estimator}\label{algorithm: PSI}
\begin{algorithmic}
\State \text{Input: primary data} $\{\bmD_i\}_{i=1}^n$, \text{secondary data} $\{\tildebmD_{mi}\}_{i=1}^n.$
\State \text{Obtain $\hatbmtheta_{m}$ using Algorithm \ref{algorithm: PEL}.}
\State \text{Obtain $\hatbmlambda_{m}$ by solving $\sum_{i=1}^n\{1+\hatbmlambda_m^T\bmg_m(\tildebmD_{mi};\hatbmtheta_m)\}^{-1} \bmg_m(\tildebmD_{mi};\hatbmtheta_m)=\bmzero$.}
\State $\hatp_{mi} \leftarrow n^{-1}\{1 + \hatbmlambda_m^T(\hatbmtheta_m)\bmg_m(\tildebmD_{mi}; \hatbmtheta_m)\}^{-1}.$
\State \text{Obtain $\hatbmbeta_{PSI}$ by solving $\sum_{i=1}^{n}\hatp_{mi}\bmf(\bmD_{i};\bmbeta_{PSI})=\bmzero$.}
\State $\text{Output: } \hatbmbeta_{PSI}.$
\end{algorithmic}
\end{algorithm}

To explore the validity of PSI, we study the asymptotic property of $\hatbmbeta_{PSI}$ in Theorem \ref{asymptotic of psi}. Without loss of generality, we write $\bmtheta_{m}^\ast=(\bmtheta_{m(1)}^{\ast T}, \bmtheta_{m(2)}^{\ast T})^T$, where $\bmtheta_{m(1)}^{\ast}$ and $\bmtheta_{m(2)}^{\ast}$ are respective zero and non-zero components of $\bmtheta_{m}^{\ast}$. Let the dimension of $\bmtheta_{m(1)}^{\ast}$ be $q_m$, and $\bmH_m$ be a matrix such that $\bmH_m\bmtheta_{m}^\ast=\bmtheta_{m(1)}^{\ast}$. 
\begin{theorem}
\label{asymptotic of psi}
Under regularity conditions listed in the Supplementary Materials, we have $\hatbmtheta_{m(1)}=\bmzero$ with probability tending to one; $\hatbmbeta_{PSI}$ is consistent to $\bmbeta_0$; $\sqrt{n}\left(\hatbmbeta_{PSI}-\bmbeta_0\right)$ converges in distribution to normal distribution $N(\bmzero, V_{PSI})$. Here 
$V_{PSI}=\bmGamma^{-1}(\bmSigma-\bmLambda_m\bmS_m\bmLambda_m^T-\bmLambda_m\bmP_m\bmLambda_m^T)\bmGamma^{-1,T}$,
where $\bmGamma=E\{\partial \bmf(\bmD;\bmbeta_0)/\partial \bmbeta^T\}$, 
$\bmSigma=E\{ \bmf(\bmD;\bmbeta_0)\linebreak\bmf^T(\bmD;\bmbeta_0)\}$, 
$\bmLambda_m=E\{ \bmf(\bmD;\bmbeta_0)\bmg_m^T(\tildebmD_{m};\bmtheta_m^{\ast})\}$,
$\bmS_m=\bmA_m^{-1}-\bmA_m^{-1}\bmB_m\bmC_m\bmB_m^T\bmA_m^{-1}$,
$\bmP_m=\bmA_m^{-1}\bmB_m\bmC_m\bmH_m^T(\bmH_m\bmC_m\bmH_m^T)^{-1}\bmH_m\bmC_m\bmB_m^T\bmA_m^{-1}$,
where
$\bmA_m=E\{ \bmg_m(\tildebmD_{m};\bmtheta_m^{\ast})\bmg_m^T(\tildebmD_{m};\bmtheta_m^{\ast})\}$,
$\bmB_m=E\{\partial \bmg_m(\tildebmD_{m};\bmtheta_m^{\ast})/\partial\bmtheta_m^T\}$,
$\bmC_m=(\bmB_m^T\bmA_m^{-1}\bmB_m)^{-1}$.
\end{theorem}

There are several implications regarding the efficiency of $\hatbmbeta_{PSI}$. First, we compare the efficiency of $\hatbmbeta_{PSI}$, $\hatbmbeta_{VIS}$, and $\hatbmbeta_{naive}$, where $\hatbmbeta_{VIS}$ is the VIS estimator proposed by \cite{chen2022improving} without adapting the penalized procedure, and $\hatbmbeta_{naive}$ is the estimator that does not borrow information from the secondary data, obtained by solving $\sum_{i=1}^{n}\bmf(\bmD_{i};\bmbeta)=\bmzero$. Note that $\bmGamma^{-1}\left(\bmSigma - \bmLambda_m \bmS_m \bmLambda_m^T \right)\bmGamma^{-1,T}$ and $\bmGamma^{-1} \bmSigma \bmGamma^{-1,T}$ are the asymptotic variances of $\hatbmbeta_{VIS}$ and $\hatbmbeta_{naive}$, respectively, with two matrices $\bmS_m$ and $\bmP_m$ being non-negative definite. The efficiency comparison among the three estimators is summarized below: in general, $\hatbmbeta_{PSI}$ is more efficient than $\hatbmbeta_{VIS}$, and $\hatbmbeta_{VIS}$ is more efficient than $\hatbmbeta_{naive}$. The superiority of $\hatbmbeta_{PSI}$ over $\hatbmbeta_{naive}$ arises from two aspects of variance reduction.
% Let the dimensions of $\bmg_m(\tildebmD_{m};\bmtheta_m)$ and $\bmtheta_{m(1)}^{\ast}$ be $r_m$ and $q_m$ respectively, and recall that the dimension of $\bmtheta_{m}$ is $p_m$ $(0\le q_m\le p_m\le r_m)$. 
The first comes from the term $\bmGamma^{-1}\bmLambda_m\bmS_m\bmLambda_m^T\bmGamma^{-1,T}$, which is non-zero and non-negative definite if the estimating function $\bmg_m(\tildebmD_{m};\bmtheta_m)$ is over-identified in its original construction. This can be achieved, for example, by forcing some components of $\hat{\bm{\theta}}_m$ to be exactly zero when it is known that the corresponding covariates have no effect on the secondary outcome. This variance reduction term is also shared by $\hatbmbeta_{VIS}$. However, it vanishes when no prior knowledge is available to construct an over-identified function.
% Note that the rank of $\bmS_m$ is $r_m-p_m$, if $\bmg_m(\tildebmD_{m};\bmtheta_m)$ is just-identified, that is $r_m=p_m$, then $\bmS_m$ becomes a zero matrix and the variance reduction vanishes. 
% discussed that $\hatbmbeta_{VIS}$ reduces to $\hatbmbeta_{naive}$ when $\bmg_m(\tildebmD_{m};\bmtheta_m)$ is just-identified.
The second comes from the term $\bmGamma^{-1}\bmLambda_m\bmP_m\bmLambda_m^T\bmGamma^{-1,T}$, which is achieved by identifying zero entries in $\bmtheta_m^\ast$ in a data-driven manner, without requiring prior knowledge. This additional variance reduction explains the superiority of $\hatbmbeta_{PSI}$ over $\hatbmbeta_{VIS}$. 
% We refer readers to Section \ref{The advantages of using PEPSI} for more discussion.
% We notice here that the rank of $\bmH_m$ is $q_m$, so the rank of $\bmP_m$ is also $q_m$. 
% This variance reduction term $\bmGamma^{-1} \bmLambda_m \bmP_m \bmLambda_m^T \bmGamma^{-1,T}$ will vanish, and $\hatbmbeta_{PSI}$ will reduce to $\hatbmbeta_{VIS}$ if there are no zero entries in $\bmtheta_m^\ast$.

The efficiency improvement of $\hatbmbeta_{PSI}$ also depends on the association between the primary and secondary data, measured by $\bmLambda_m$. A stronger association leads to greater variance reduction, whereas if the data are independent, $\bmLambda_m$ becomes zero and the secondary data do not contribute to the primary data analysis. Note that $\bmLambda_m$ depends not only on the association between $Y$ and $\tildeY_m$ but also on the relationship between $\bmX$ and $\tildebmX_m$. For example, when both the primary and secondary outcomes are cross-sectional and follow generalized linear models, $\bmLambda_m = E\{\text{cov}(Y, \tildeY_m|\bmX, \tildebmX_m)\bmX \tildebmX_m^T\}$. Therefore, it is recommended to always include $\bmX$ as part of $\tildebmX_m$, as this generally enhances integration performance. 

Moreover, the efficiency superiority of $\hatbmbeta_{PSI}$ is robust to misspecification of the secondary working model. Since inference on the secondary data is not of direct interest, $\bmg_m(\tildebmD_{m};\bmtheta_m)$ dose not need to be correctly specified, nor does $\bmtheta_m^\ast$ need to represent the true parameter generating the secondary data. Instead, it is sufficient that there exists some $\bmtheta_m^\ast$ satisfying $E\{\bmg_m(\tildebmD_{m}; \bmtheta_m^\ast)\} = \bmzero$. This robustness property is demonstrated through numerical experiments in Section \ref{Simulation - Case 1}.

Finally, if all coefficients in $\bmtheta_{m}^\ast$ are non-zero, which may happen with a single secondary data with a limited set of covariates, and no prior knowledge is available to guide the construction of the over-identified function, the PSI estimator will not yield efficiency gains. However, this limitation can be mitigated by using multiple secondary outcomes, where some coefficients in $\bmtheta_m^\ast$ may be zero for certain $m$, but not for all $m$. This naturally motivates the need for a new method to integrating multiple secondary outcomes, as described in the following section.

\subsection{Multiple Secondary Data}\label{Multiple Secondary Data}
The PSI framework provides the essential foundation for the PEPSI framework, which addresses a key research challenge: how to effectively leverage multiple secondary data sources to enhance statistical performance in the primary analysis. In our data application, we have $22$ blood biochemistry markers and $3$ urine biomarkers in addition to the primary outcome, all of which are highly associated with liver health. Integrating information from these multiple secondary outcomes can significantly improve the primary outcome analysis. 

Suppose there are $M$ secondary data $\{\tildebmD_m\}_{m=1}^M$, and we aim to integrate all of them into the primary data analysis. 
% A natural question is, how to integrate extra information from multiple datasets and fuse it to the main model inference efficiently and robustly? 
A natural approach is to stack all working estimating functions $\{\bmg_m(\tildebmD_m,\bmtheta_m)\}_{m=1}^M$ and form a pooled estimating function $\bmG(\tildebmD;\bmTheta) = (\bmg_1^T(\tildebmD_{1};\bmtheta_1), \dots, \bmg_M^T(\tildebmD_{M}; \bmtheta_M))^T,$ where $\tildebmD = \{\tildebmD_1, \dots, \tildebmD_M\}$ and $\bmTheta = (\bmtheta_1^T, \dots, \bmtheta_M^T)^T$ are the pooled data and pooled parameter vector, respectively. One could then construct a PSI estimator, as described in Section \ref{Single Secondary Data}, with $\bmg_m(\tildebmD_m, \bmtheta_m)$ replaced by $\bmG(\tildebmD; \bmTheta)$. However, this approach becomes computationally challenging due to the high dimensions of both $\bmG(\tildebmD; \bmTheta)$ and $\bmTheta$, as well as the strong  association and potential co-linearity among functions in $\bmG(\tildebmD; \bmTheta)$. These features hinder the broad applicability of joint estimation. {For example, suppose we have $25$ secondary outcomes that may exhibit substantial correlation. Even if each secondary outcome model contains only $4$ parameters, this still yields $100$ functions in the $\bmG(\tildebmD; \bmTheta)$ vector and $100$ parameters in $\bmTheta$.} The dimensionality grows much larger when the number of secondary outcomes or parameters in each model increases. Rather than solving all parameters in $\bmTheta$ simultaneously, we adopt a divide-and-conquer strategy. Specifically, we obtain estimators $\{\hatbmtheta_m\}_{m=1}^M$ from each secondary data separately, as described in Section \ref{Single Secondary Data}, and then pool them to form the enlarged estimator vector $\hatbmTheta$. This alternative avoids the computationally intensive task of solving the entire parameter space within a single penalized and constrained optimization step, thereby substantially reducing the computational load.
% One advantage of doing so is that, we can reduce the dimensions by filtering out secondary data for which the estimating functions are just-identified, and the parameter estimates do not have zero entry, as they do not carry extra information based on our discussion in Section \ref{Single Secondary Data}. 

Despite the computational relief gained from estimating $\bmTheta$, the dimension of the enlarged vector $\bmG(\tildebmD; \hatbmTheta)$ remains high. More importantly, because the plug-in estimator $\hatbmTheta$ is obtained via the divide-and-conquer approach rather than by directly solving the pooled estimating equation, it becomes unclear how the resulting $\bmG(\tildebmD; \hatbmTheta)$ would contribute to the primary data analysis. Consequently, the weights defined in Section \ref{Single Secondary Data} may not be directly applicable. 

One strategy for integrating multiple secondary outcomes is to design an Averaging scheme that takes a linear combination of $\{\hatp_{mi}\}_{m=1}^M$, obtained by PSI across the secondary outcomes. Specifically, the Averaging scheme estimator $\hatbmbeta_{avg}$ is defined as the solution to the estimating equation $\sum_{i=1}^{n}\left(\sum_{m=1}^M w_m\hatp_{mi}\right)\bmf(\bmD_{i};\bmbeta)=\bmzero$,
where the hyperparameters $\{w_m\}_{m=1}^M$, satisfying $\sum_{m=1}^M w_m=1$, are the weights assigned for each secondary outcome and can be determined via data-driven optimization \citep{chen2023efficient}. Details of this method are described in Section 1 in the Supplementary Materials. 
% We discuss the asymptotic results of $\hatbmbeta_{avg}$ and the ways to determine $\{w_m\}_{m=1}^M$ in  
We note that $\hatbmbeta_{avg}$ is new in that it combines the strengths of the proposed PSI and the MinBo approach \citep{chen2023efficient}. 
% Strictly following the averaging scheme from MinBo will yield little efficiency gain if there is no prior knowledge to guide the construction of the over-identified function. 
However, a potential limitation is that it may not optimally exploit the information from the secondary outcomes. In what follows, we introduce another new perspective, PEPSI, which constructs weights by adapting dimensionality reduction and projection techniques to address this limitation.

To motivate the construction of new weights, we examine the  
% For each secondary data $\tildebmD_m$, given a working estimating function $\bmg_m(\tildebmD_{m};\bmtheta_m)$, since we do not know the true value $\bmtheta_m^\ast$, we use an estimate $\hatbmtheta_m$ to approximate it, and some information is lost due to the estimation. This is reflected in the 
asymptotic formula 
$\text{var}\left\{n^{-1/2}\sum_{i=1}^n\bmg_m(\tildebmD_{mi};\hatbmtheta_m)\right\}=\bmR_m \bmA_m\bmR_m^T+o_p(1),$
where $\bmR_m=\bmA_m(\bmS_m+\bmP_m)$ with $\bmA_m$, $\bmS_m$, and $\bmP_m$ defined in Theorem \ref{asymptotic of psi}. We notice here that the rank of $\bmR_m$ is $r_m-(p_m-q_m)$, so is the rank of $\bmR_m \bmA_m\bmR_m^T$, where $r_m$, $p_m$ and $q_m$ are the dimensions of $\bmg_m(\tildebmD_{m};\bmtheta_m)$, $\bmtheta_m$ and $\bmtheta_{m(1)}^\ast$ (defined in Section \ref{Single Secondary Data}), respectively. This rank can be understood in terms of two components: $r_m$ and $p_m - q_m$. The first component, $r_m$, reflects the total degrees of freedom (DoF) in $\bmg_m(\tildebmD_{m}; \bmtheta_m^\ast)$ when the parameters in $\bmtheta^\ast_m$ are fully known. The second component, $(p_m - q_m)$, represents the loss of DoF due to estimating the non-zero components of $\bmtheta^\ast_m$. Thus, the difference between these two components indicates the remaining DoF that can contribute to the primary analysis. In other words, there exists a low-dimensional representation within $\bmG(\tildebmD; \hatbmTheta)$, referred to as the ``informative component," which plays an essential role in data integration. This observation motivates us to locate the informative component that benefits primary model inference by reducing the dimensionality of $\bmG(\tildebmD; \hatbmTheta)$. To this end, we propose an extended PSI framework, named PEPSI, which involves two steps for weight construction.

\textbf{Step 1 (dimension reduction)}: We apply principal component analysis (PCA) to $\hatbmR\bmG(\tildebmD_i;\hatbmTheta)$, $i=1,2,\dots,n$, where $\bmR=\text{diag}(\bmR_1,\dots,\bmR_M)$, and $\hatbmR$ is the sample estimate of $\bmR$. Let $\hatK$ be the rank of the sample variance matrix, approximated by $\sum_{m=1}^M r_m-(p_m-\hatq_m)$ based on the above discussion, where $\hatq_m$ is the number of zero entries in $\hatbmtheta_{m(1)}$. Let $\hatbmPhi$ and $\hatbmU$ be the diagonal matrix of the first $\hatK$ eigenvalues and the corresponding rotation matrix of eigenvectors obtained from PCA.

\textbf{Step 2 (weight construction)}: We then construct the weights using the following expression to ensure valid data integration: $\hatp_{i}=n^{-1}\{1-n^{-1}\bmG^T(\tildebmD_{i};\hatbmTheta)\hatbmR^T\hatbmU\hatbmPhi^{-1}\hatbmU^T\hatbmR \linebreak\sum_{i=1}^n\bmG(\tildebmD_{i};\hatbmTheta)\}.$ The intuition behind constructing these weights is to project $\bmf(\bmD;\bmbeta)$ onto the orthogonal complement of linear space spanned by $\bmG(\tildebmD;\hatbmTheta)$, thereby minimizing the estimator variance (see Corollary \ref{optimality of pepsi} and its proof in the Supplementary Materials for technical details). After weight construction, the PEPSI estimator $\hatbmbeta_{PEPSI}$ is obtained by solving the re-weighted estimating equation: $\sum_{i=1}^{n}\hatp_{i}\bmf(\bmD_{i};\bmbeta)=\bmzero$.

The pseudo algorithm for implementing PEPSI is provided in Algorithms \ref{algorithm: PEL} and \ref{algorithm: PEPSI}, and the asymptotic properties of $\hatbmbeta_{PEPSI}$ are summarized in Theorem \ref{asymptotic pepsi}.

\begin{algorithm}[ht]
\caption{Algorithm for obtaining the PEPSI estimator}\label{algorithm: PEPSI}
\begin{algorithmic}
\State \text{Input: primary data} $\{\bmD_i\}_{i=1}^n$, \text{secondary data} $\{\tildebmD_{i}\}_{i=1}^n.$
\State \text{Obtain $\{\hatbmtheta_m\}_{m=1}^M$ using Algorithm \ref{algorithm: PEL} and set $\hatbmTheta \leftarrow (\hatbmtheta_1^T, \dots, \hatbmtheta_M^T)^T$.}
\State \text{Obtain $\hatbmPhi$ and $\hatbmU$ by applying PCA on $\hatbmR\bmG(\tildebmD_i;\hatbmTheta)$.}
\State $\hatp_{i} \leftarrow n^{-1}\left\{1-n^{-1}\bmG^T(\tildebmD_{i};\hatbmTheta)\hatbmR^T\hatbmU\hatbmPhi^{-1}\hatbmU^T\hatbmR\sum_{i=1}^n\bmG(\tildebmD_{i};\hatbmTheta)\right\}.$
\State \text{Obtain $\hatbmbeta_{PEPSI}$ by solving $\sum_{i=1}^{n}\hatp_{i}\bmf(\bmD_{i};\bmbeta_{PEPSI})=\bmzero$.}
\State $\text{Output: } \hatbmbeta_{PEPSI}.$
\end{algorithmic}
\end{algorithm}

\begin{theorem}\label{asymptotic pepsi}
Under mild conditions listed in the Supplementary Materials,  $\hatbmbeta_{PEPSI}$ is consistent to $\bmbeta_0$, and $\sqrt{n}\left(\hatbmbeta_{PEPSI}-\bmbeta_0\right)$ converges in distribution to $N(\bmzero, V_{PEPSI})$. Here, $V_{PEPSI}=\bmGamma^{-1}\left(\bmSigma-\bmLambda\bmR^T\bmU\bmPhi^{-1}\bmU^T\bmR\bmLambda^T \right)\bmGamma^{-1,T}$,
where $\bmGamma$ and $\bmSigma$ are defined in Theorem 1; $\bmLambda=E\{ \bmf(\bmD;\bmbeta_0)\bmG^T(\tildebmD;\bmTheta^{\ast})\}$;
$\bmPhi$ and $\bmU$ are the diagonal matrix and rotation matrix containing the first $K$ eigenvalues and eigenvectors of $\bmW = \bmR E\left\{\bmG(\tildebmD;\bmTheta^\ast)\bmG^T(\tildebmD;\bmTheta^\ast)\right\}\bmR^T$, respectively; $\bmTheta^\ast$ is the pooled true value vector satisfying $E\left\{\bmG(\tildebmD;\bmTheta^\ast)\right\}=\bmzero$ and $K$ is the rank of $\bmW$.
\end{theorem}

Theorem \ref{asymptotic pepsi} shows that the PEPSI framework effectively reduces the variance in estimating $\bmbeta$. A key insight is that identifying zero coefficients is essential for leveraging additional information, and incorporating multiple secondary outcomes, rather than relying on a single one, is beneficial, as it increases the likelihood of encountering zero coefficients across various secondary models. In other words, as long as some coefficients in $\bmtheta_m^\ast$ are zero for at least one $m$, and more importantly, without requiring prior knowledge of which secondary outcomes contain zero coefficients, the PEPSI estimator is more likely to achieve improved efficiency.

\begin{corollary}
\label{optimality of pepsi}
Given a collection of working estimating functions $\{\bmg_m(\tildebmD_m;\bmtheta_m)\}_{m=1}^M$ and penalized empirical likelihood estimates $\{\hatbmtheta_m\}_{m=1}^M$ from \eqref{penalized loss}, $\hatbmbeta_{PEPSI}$ is the most efficient estimator among the following class of estimating equations
$$\left\{\frac{1}{n}\sum_{i=1}^{n}\left\{\bmf(\bmD_{i};\bmbeta)+\sum_{m=1}^M\bmF_m\bmg_m(\tildebmD_{mi};\hatbmtheta_m)\right\}=o_p(n^{-1/2}) \right\},$$
where $\{\bmF_m\}_{m=1}^M$ are any arbitrary conformable matrices.
\end{corollary}

In addition to investigating the asymptotic distributions, we have examined other theoretical properties of the PSI/PEPSI estimators (see Section 1 of the Supplementary Materials). Corollary \ref{optimality of pepsi} shows the optimality of PEPSI estimator within a specific class of estimators, including $\hatbmbeta_{avg}$. This analysis highlights the estimator's ability to achieve the smallest variance and thus demonstrates that PEPSI effectively leverages secondary outcomes to maximize efficiency. It is also worth noting that $\hatbmbeta_{PSI}$ and $\hatbmbeta_{PEPSI}$ are equivalent when there is only a single secondary outcome. Hence, $\hatbmbeta_{PEPSI}$ can be viewed as a generalization of $\hatbmbeta_{PSI}$.

We conclude the method section with a discussion on secondary outcome integration within the PEPSI framework. Although the proposed framework incorporates both primary and secondary outcomes into the analysis, our theoretical results show that, under mild regularity conditions, the data integration process does not alter the underlying effect size, $\bmbeta_0$, or target population defined by the original study design. Instead, the framework offers substantial gains in estimation efficiency without compromising validity. To maximize the utility of our proposed methods, careful selection of secondary outcomes prior to analysis is essential. We recommend that, at the study design stage, investigators identify secondary outcomes plausibly associated with the primary outcome, guided by prior literature or clinical expertise. During the analysis stage, a manageable number of secondary outcomes should be selected based on the sample size. A practical strategy is to prioritize secondary outcomes according to their strength of association with the primary outcome, for example, using Kendall’s tau correlation coefficient. Additional application guidance is provided in Section \ref{Discussion}.

\section{Simulation}\label{simulation}
We examined the numerical performance of our proposal under two main cases. In the first case, we considered a continuous primary outcome and one continuous secondary outcome. In the second case, we considered a continuous primary outcome and multiple continuous secondary outcomes. Besides, we examined extra cases: 1) the VIS estimator constructed with and without prior knowledge of zero coefficients; 2) a comparison of the PSI and PEPSI estimators to assess their equivalence when only one secondary outcome is present; 3) a binary primary outcome; 4) the secondary outcomes were mildly associated with the primary outcome; 5) the secondary outcomes with a lower signal-to-noise ratio; 6) prediction of the primary outcome from covariates and secondary outcomes using a machine learning model; and 7) small sample sizes. Details are presented in Section 3 of the Supplementary Materials. Ordinary least squares (OLS) regression and logistic regression were used for continuous and binary outcomes, respectively. For each case, we ran $10,000$ Monte Carlo replications and reported the estimation bias, Monte Carlo standard deviation (MCSD), estimated standard error (SE), coverage probability of the $95\%$ confidence interval (CP), and relative efficiency (RE; i.e., variance ratio compared with the naive estimator; larger than $1$ implies efficiency gain).

\subsection{Case 1: one continuous primary outcome and one continuous secondary outcome}
\label{Simulation - Case 1}
We drew $n$ i.i.d samples from the following model: 
$$\begin{pmatrix}
    Y\\
    \tildeY_1
\end{pmatrix}= \begin{pmatrix}
    \beta_0\\
    0
\end{pmatrix} + \begin{pmatrix}
    \bmbeta^T\\
    \bmtheta_1^T
\end{pmatrix}\bmX + \bmepsilon=
\begin{pmatrix}
    1\\
    0
\end{pmatrix}+
\begin{pmatrix}
    1 & 1 & 1 & 1\\
    1 & 1 & 0 & 0
\end{pmatrix} \bmX + \bmepsilon,
$$
where $Y$ and $\tildeY_1$ are the primary and secondary outcomes, respectively; $\beta_0$ is the intercept, while $\bmbeta=(\beta_1,...,\beta_4)^T$ and $\bmtheta_1=(\theta_{11},...,\theta_{14})^T$ are the coefficient vectors. Some elements of $\bmtheta_1$ were set to zero to facilitate data integration. As described in Section \ref{Single Secondary Data}, 
null effects from certain covariates on outcomes are common in practice. However, it is important to note that our proposed methods do not require any prior knowledge to identify null-effect covariates, nor do they assume the primary parameters in $\bmbeta$ to be zero or non-zero. For illustration, we considered the case with $\bmbeta = (1, \ldots, 1)^T$. The covariate vector $\bmX=(X_1,...,X_4)^T$ followed a multivariate normal distribution with mean 0, variance 1, and AR(1) correlation with parameter $0.5$. The random noise term $\bmepsilon$ followed a bivariate normal distribution with mean 0, variance 1, and correlation coefficient $\rho$. We considered two sample sizes, $n=300$ and 600, and two correlation levels between $Y$ and $\tildeY_1$, $\rho=0.5$ or $0.8$. 
% We estimate $(\beta_0,\bmbeta^T)^T$ using the PSI estimator $\hatbmbeta_{PSI}$. 
For each configuration, we checked both the correctly specified model and a misspecified working model ($X_1$ was excluded from the working model) for the secondary analysis.

\begin{table}[ht]
\centering
\caption{Simulation results of Case 1. All results except RE are multiplied by 100.}
\label{table1 - single secondary outcome}
\resizebox{0.9\textwidth}{!}{%
\begin{tabular}{@{}ccccccccccccc@{}}
\toprule
 &  &  & \multicolumn{5}{c}{Correctly specified model} & \multicolumn{5}{c}{Misspecified model} \\
 &  &  & Bias & MCSD & SE & CP & RE & Bias & MCSD & SE & CP & RE \\ \midrule
\multirow{11}{*}{n=300} & \multirow{5}{*}{$\rho$=0.5} & $\beta_0$ & 0.0 & 5.9 & 5.7 & 95 & 1.0 & 0.0 & 5.9 & 5.7 & 95 & 1.0 \\
 &  & $\beta_1$ & 0.0 & 6.7 & 6.6 & 95 & 1.0 & 0.2 & 6.7 & 6.6 & 95 & 1.0 \\
 &  & $\beta_2$ & 0.1 & 7.5 & 7.2 & 94 & 1.0 & 0.0 & 7.6 & 7.3 & 94 & 1.0 \\
 &  & $\beta_3$ & -0.1 & 6.7 & 6.4 & 94 & 1.3 & 0.0 & 7.0 & 6.9 & 95 & 1.2 \\
 &  & $\beta_4$ & 0.1 & 5.9 & 5.8 & 94 & 1.3 & 0.1 & 6.3 & 6.1 & 94 & 1.1 \\
 &  &  &  &  &  &  &  &  &  &  &  &  \\
 & \multirow{5}{*}{$\rho$=0.8} & $\beta_0$ & 0.0 & 5.9 & 5.7 & 94 & 1.0 & 0.0 & 5.9 & 5.7 & 94 & 1.0 \\
 &  & $\beta_1$ & 0.0 & 6.7 & 6.6 & 95 & 1.0 & 0.2 & 6.7 & 6.6 & 95 & 1.0 \\
 &  & $\beta_2$ & 0.1 & 7.2 & 6.9 & 94 & 1.1 & -0.1 & 7.4 & 7.1 & 94 & 1.1 \\
 &  & $\beta_3$ & -0.1 & 4.9 & 4.5 & 94 & 2.4 & 0.0 & 6.1 & 5.9 & 95 & 1.5 \\
 &  & $\beta_4$ & 0.1 & 4.4 & 4.1 & 94 & 2.3 & 0.1 & 5.5 & 5.3 & 94 & 1.5 \\ \midrule
\multirow{11}{*}{n=600} & \multirow{5}{*}{$\rho$=0.5} & $\beta_0$ & 0.0 & 4.1 & 4.1 & 95 & 1.0 & 0.0 & 4.1 & 4.1 & 95 & 1.0 \\
 &  & $\beta_1$ & 0.0 & 4.7 & 4.7 & 94 & 1.0 & 0.1 & 4.7 & 4.7 & 94 & 1.0 \\
 &  & $\beta_2$ & 0.0 & 5.2 & 5.1 & 95 & 1.1 & -0.1 & 5.2 & 5.2 & 95 & 1.0 \\
 &  & $\beta_3$ & -0.1 & 4.6 & 4.6 & 94 & 1.3 & -0.1 & 4.9 & 4.9 & 95 & 1.2 \\
 &  & $\beta_4$ & 0.1 & 4.1 & 4.1 & 95 & 1.3 & 0.0 & 4.4 & 4.4 & 94 & 1.1 \\
 &  &  &  &  &  &  &  &  &  &  &  &  \\
 & \multirow{5}{*}{$\rho$=0.8} & $\beta_0$ & 0.0 & 4.1 & 4.1 & 95 & 1.0 & 0.0 & 4.1 & 4.1 & 95 & 1.0 \\
 &  & $\beta_1$ & 0.0 & 4.7 & 4.7 & 95 & 1.0 & 0.1 & 4.8 & 4.7 & 95 & 1.0 \\
 &  & $\beta_2$ & -0.1 & 5.0 & 4.9 & 94 & 1.1 & -0.1 & 5.2 & 5.1 & 95 & 1.1 \\
 &  & $\beta_3$ & -0.1 & 3.3 & 3.2 & 94 & 2.6 & -0.1 & 4.3 & 4.2 & 95 & 1.5 \\
 &  & $\beta_4$ & 0.1 & 3.0 & 2.9 & 94 & 2.5 & 0.0 & 3.9 & 3.8 & 94 & 1.5 \\ \bottomrule
\end{tabular}%
}
\begin{tablenotes}[flushleft]
\item \footnotesize Abbreviations: MCSD, Monte Carlo standard deviation; SE, estimated standard error; RE, relative efficiency over the naive estimator; CP, 95\% coverage probability.
\end{tablenotes}
\end{table}

The results for $\hatbmbeta_{PSI}$ are summarized in Table \ref{table1 - single secondary outcome}. Overall, the bias was negligible, the estimated SE was close to the MCSD, and the CP was near the nominal level, confirming that $\hatbmbeta_{PSI}$ is consistent and that the asymptotic theory works well in finite samples. Compared with the naive estimator, $\hatbmbeta_{PSI}$ was more efficient, i.e., smaller variance, especially for the coefficients ($\beta_{3}$ and $\beta_{4}$) whose corresponding coefficients in the secondary analysis were $0$ ($\theta_{13}$ and $\theta_{14}$). The efficiency gain became more prominent as the correlation $\rho$ increased, which was expected since the secondary outcome provided more information for the primary outcome. 
% Note that $\hatbmbeta_{PSI}$ also becomes more efficient with increasing sample sizes. This is partially due to the fact that the estimates of $\theta_{13}$ and $\theta_{14}$ are more concentrated around $0$ with a higher probability. 
Moreover, $\hatbmbeta_{PSI}$ remained unbiased and more efficient even under a misspecified working model, confirming the robustness of our proposal. More evaluations for $\hatbmbeta_{naive}$ and $\hatbmbeta_{VIS}$ with and without prior knowledge about zero coefficients are presented in Table S1 in the Supplementary Materials.

\subsection{Case 2: one continuous primary outcome and multiple continuous secondary outcomes}

Next we generalized Case 1 by considering in total $50$ secondary outcomes. We adopted the same data-generating process for the covariate $\bmX$ and the primary outcome $Y$ as in Case 1. In addition to the secondary outcome $\tildeY_1$ with $\rho=0.8$ in Case 1, we generated $49$ more secondary outcomes. The model for the $50$ secondary outcomes is $(\tildeY_1,\dots,\tildeY_{50})^T= \bmTheta^T \bmX + \bmepsilon$, where $\bmTheta$ is the coefficient matrix, with details specified in Section 3 of the Supplementary Materials. Briefly, the first $10$ secondary outcomes had some zero entries in their coefficient vectors, while the remaining $40$ outcomes had all non-zero entries. This setup mimics the real-world situation where only a subset of secondary outcomes have null effects for certain covariates. The random noise term $\bmepsilon$ followed a multivariate normal distribution with mean $0$, variance $1$, and constant correlation coefficient $0.5$. The residual correlations between the primary outcome $Y$ and the secondary outcomes $\{\tildeY_1,\dots,\tildeY_{50}\}$ were
$0.8, 0.8, 0.5, 0.5, \dots, 0.5$, respectively. We considered two sample sizes, $n=300$ or $600$, and two scenarios for integrating the secondary outcomes: (1) integrating $10$ secondary outcomes $\{\tildeY_1, \tildeY_{7}, \tildeY_{8},\dots,\tildeY_{15}\}$; and (2) integrating 50 secondary outcomes $\{\tildeY_1,\dots,\tildeY_{50}\}$. For each case, we evaluated and compared the performance of $\hatbmbeta_{PEPSI}$ and $\hatbmbeta_{avg}$. In particular, regarding $\hatbmbeta_{avg}$, $\{w_m\}_{m=1}^M$ were determined by maximizing the sum of proportion of sample variance reduction (see Section 1 of the Supplementary Materials).

\begin{table}[ht]
\centering
\caption{Simulation results of Case 2. All results except RE are multiplied by 100.}
\label{table2 - multiple secondary outcomes}
\resizebox{0.9\textwidth}{!}{%
\begin{tabular}{@{}ccccccccccccc@{}}
\toprule
 &  &  & \multicolumn{5}{c}{PEPSI} & \multicolumn{5}{c}{Averaging scheme} \\
 &  &  & Bias & MCSD & SE & CP & RE & Bias & MCSD & SE & CP & RE \\ \midrule
\multirow{11}{*}{n=300} & \multirow{5}{*}{\begin{tabular}[c]{@{}c@{}}Integrating\\ 10 outcomes \end{tabular}} & $\beta_0$ & 0.0 & 5.9 & 5.7 & 94 & 1.0 & 0.0 & 5.9 & 5.7 & 94 & 1.0 \\
 &  & $\beta_1$ & 0.0 & 5.7 & 5.3 & 93 & 1.4 & 0.0 & 6.5 & 6.3 & 94 & 1.1 \\
 &  & $\beta_2$ & 0.0 & 6.1 & 5.6 & 93 & 1.6 & -0.1 & 6.9 & 6.6 & 94 & 1.2 \\
 &  & $\beta_3$ & 0.0 & 4.7 & 4.3 & 93 & 2.6 & 0.0 & 4.9 & 4.6 & 94 & 2.3 \\
 &  & $\beta_4$ & -0.1 & 4.2 & 3.9 & 93 & 2.5 & -0.1 & 4.4 & 4.1 & 94 & 2.3 \\
 &  &  &  &  &  &  &  &  &  &  &  &  \\
 & \multirow{5}{*}{\begin{tabular}[c]{@{}c@{}}Integrating\\ 50 outcomes\end{tabular}} & $\beta_0$ & 0.1 & 6.0 & 5.6 & 93 & 0.9 & 0.1 & 5.8 & 5.7 & 94 & 1.0 \\
 &  & $\beta_1$ & 0.0 & 4.3 & 3.7 & 91 & 2.5 & 0.0 & 5.1 & 4.8 & 93 & 1.8 \\
 &  & $\beta_2$ & 0.0 & 4.5 & 4.0 & 91 & 2.7 & 0.0 & 5.3 & 5.0 & 94 & 2.0 \\
 &  & $\beta_3$ & 0.0 & 4.6 & 4.0 & 91 & 2.7 & -0.1 & 5.4 & 5.0 & 93 & 2.0 \\
 &  & $\beta_4$ & 0.0 & 4.3 & 3.7 & 91 & 2.5 & 0.0 & 5.1 & 4.8 & 94 & 1.8 \\ \midrule
\multirow{11}{*}{n=600} & \multirow{5}{*}{\begin{tabular}[c]{@{}c@{}}Integrating\\ 10 outcomes\end{tabular}} & $\beta_0$ & 0.0 & 4.1 & 4.0 & 95 & 1.0 & 0.0 & 4.1 & 4.1 & 95 & 1.0 \\
 &  & $\beta_1$ & 0.0 & 3.9 & 3.8 & 94 & 1.4 & 0.0 & 4.5 & 4.5 & 95 & 1.1 \\
 &  & $\beta_2$ & 0.1 & 4.2 & 4.0 & 94 & 1.6 & 0.1 & 4.8 & 4.7 & 95 & 1.2 \\
 &  & $\beta_3$ & 0.0 & 3.2 & 3.1 & 94 & 2.7 & 0.0 & 3.4 & 3.3 & 94 & 2.4 \\
 &  & $\beta_4$ & 0.0 & 2.9 & 2.7 & 94 & 2.7 & 0.0 & 3.0 & 2.9 & 94 & 2.4 \\
 &  &  &  &  &  &  &  &  &  &  &  &  \\
 & \multirow{5}{*}{\begin{tabular}[c]{@{}c@{}}Integrating\\ 50 outcomes\end{tabular}} & $\beta_0$ & 0.0 & 4.2 & 4.0 & 94 & 1.0 & 0.0 & 4.1 & 4.1 & 95 & 1.0 \\
 &  & $\beta_1$ & 0.0 & 2.9 & 2.7 & 93 & 2.7 & 0.0 & 3.5 & 3.4 & 94 & 1.8 \\
 &  & $\beta_2$ & 0.0 & 3.1 & 2.8 & 93 & 3.1 & 0.0 & 3.8 & 3.6 & 94 & 2.1 \\
 &  & $\beta_3$ & 0.0 & 3.1 & 2.8 & 93 & 3.1 & 0.0 & 3.7 & 3.6 & 94 & 2.1 \\
 &  & $\beta_4$ & 0.1 & 2.9 & 2.7 & 93 & 2.7 & 0.1 & 3.5 & 3.4 & 94 & 1.8 \\ \bottomrule
\end{tabular}%
}
\begin{tablenotes}[flushleft]
\item \footnotesize Abbreviations: MCSD, Monte Carlo standard deviation; SE, estimated standard error; RE, relative efficiency over the naive estimator; CP, 95\% coverage probability.
\end{tablenotes}
\end{table}

The results are summarized in Table \ref{table2 - multiple secondary outcomes}. Both $\hatbmbeta_{PEPSI}$ and $\hatbmbeta_{avg}$ had negligible bias. 
% As for the asymptotic variance approximation, its performance is slightly poor when the sample size is $300$ and we integrate 50 secondary outcomes, which is not surprising as we are handling a high-dimensional problem in this circumstance. 
As the sample size increased, the SE and CP approached the MCSD and the nominal $95\%$ level, respectively, validating our asymptotic theory. 
% Next, we investigate the relationship between the number of secondary outcomes and efficiency gain. Note that we have evaluated the performance of $\hatbmbeta_{PSI}$ integrating $\tildeY_1$ ($\rho=0.8$) in Case 1, and $\hatbmbeta_{PSI}$ and $\hatbmbeta_{PEPSI}$ are equivalent when integrating one secondary outcome. 
Moreover, $\hatbmbeta_{PEPSI}$ became more efficient as the number of secondary outcomes increases and showed a greater efficient gain compared with $\hatbmbeta_{avg}$.  Notably, the RE of $\hatbmbeta_{avg}$ did not necessarily improve for certain parameters (e.g. $\beta_3$ and $\beta_4$) when the number of secondary outcomes increased from $10$ to $50$. This is expected because the Averaging scheme makes some compromises among secondary outcomes, sacrificing some parameter-specific efficiency to maximize the overall proportion of sample variance reduction \citep{chen2023efficient}, whereas the PEPSI integration scheme makes better use of secondary data (see Corollary \ref{optimality of pepsi}). Similar conclusions were observed for all other cases with moderate sample sizes, as outlined in the Supplementary Materials (see Tables S3 - S7 for cases 3 - 7). Besides, estimators $\hatbmbeta_{PSI}$ and $\hatbmbeta_{PEPSI}$ performed almost identically when only one secondary outcome was available (Table S2), further validating our theoretical findings.

\begin{figure}[ht]
    \centering\includegraphics[width=.6\linewidth]{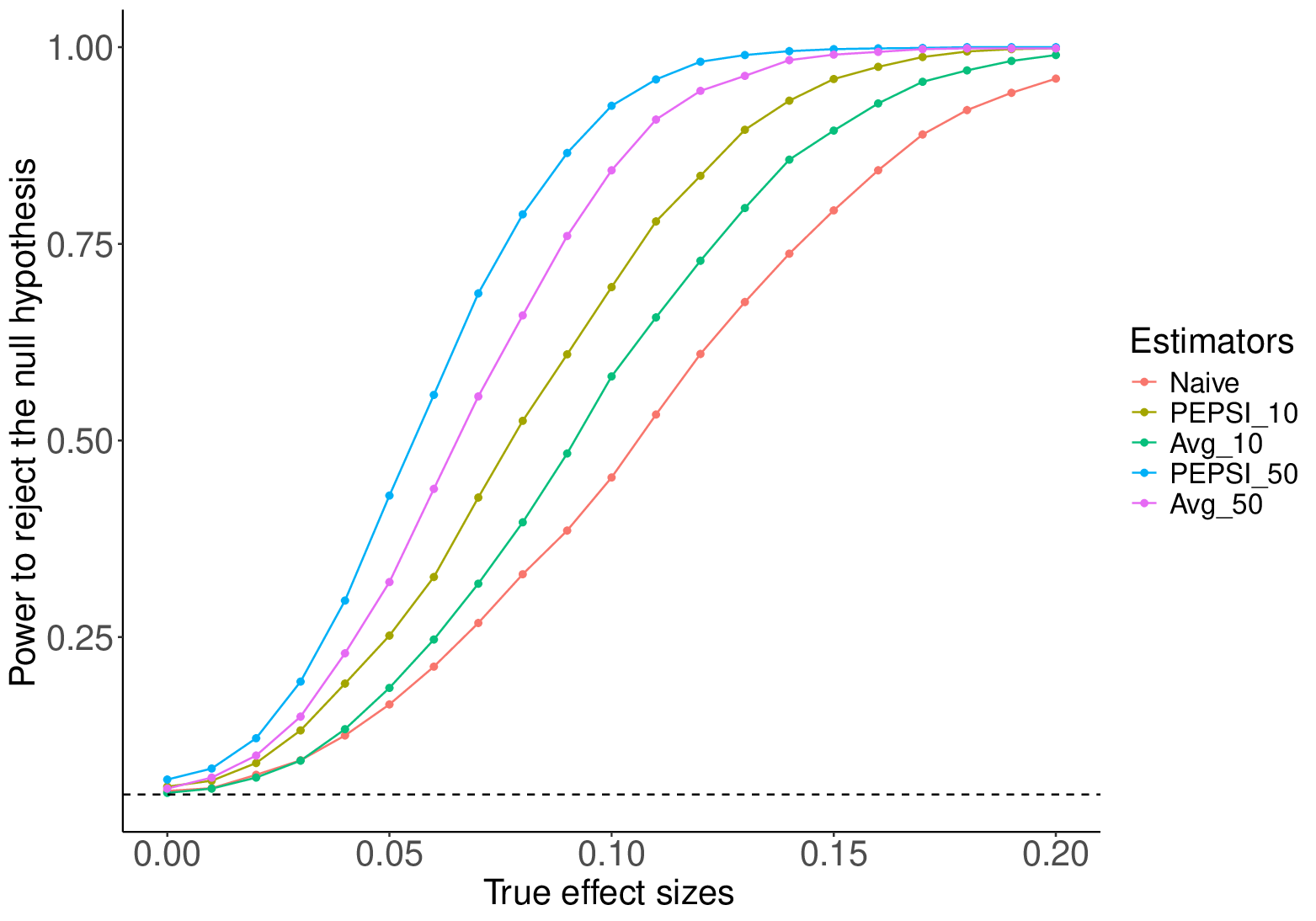}
    \caption{Statistical power evaluation of various estimators for rejecting the null hypothesis $H_0:\beta_2=0$ under different true values of $\beta_2$. Naive: the standard estimator without data integration; PEPSI\_10 (50) and Avg\_10 (50): the PEPSI and  Averaging scheme estimators integrating 10 (50) secondary outcomes, respectively.}
    \label{fig_power}
\end{figure}

To evaluate the benefit of data integration in hypothesis testing, we considered a modified Case 2 by varying the effect size $\beta_2$ from $0$ to $0.2$. Using the asymptotic normality of $\hatbmbeta_{naive}$, $\hatbmbeta_{PEPSI}$, and $\hatbmbeta_{avg}$, we constructed Wald tests and compared their rejection rates for the null hypothesis $H_0:\beta_2=0$ with a $0.05$ significance levels. The results are summarized in Figure \ref{fig_power}. Notable, both $\hatbmbeta_{PEPSI}$ and $\hatbmbeta_{avg}$ maintained type I error rates near $0.05$ under the null, while achieving greater power than $\hatbmbeta_{naive}$ under the alternative, especially as more secondary outcomes were integrated. Overall, $\hatbmbeta_{PEPSI}$ was more powerful than $\hatbmbeta_{avg}$ with moderate sample sizes.

\begin{figure}[ht]
    \centering
    \includegraphics[width= \linewidth]{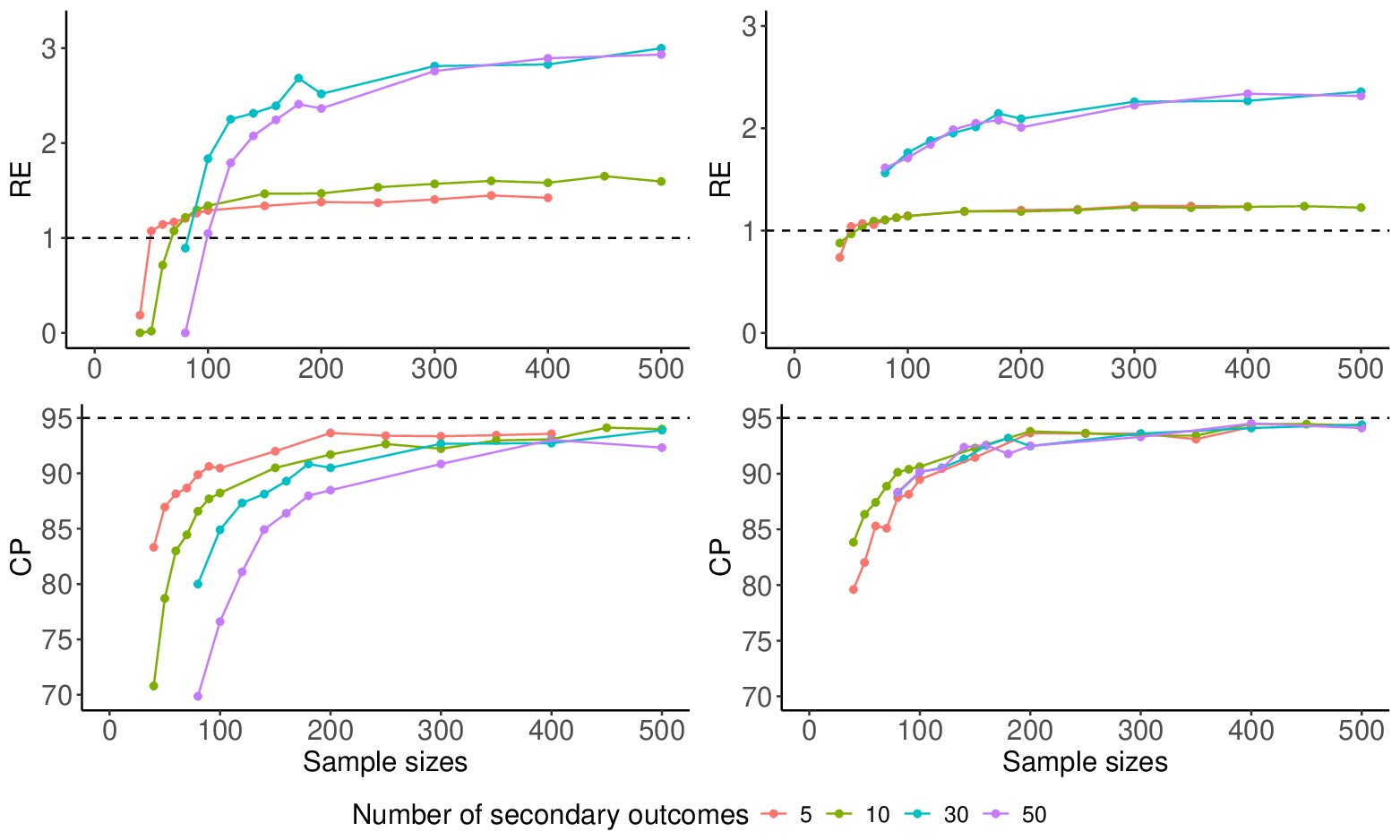}
    \caption{Relative efficiency (RE) and $95\%$ coverage probabilities (CP) of the PEPSI estimator (left) and the Averaging scheme estimator (right) for estimating $\beta_2$ across varying sample sizes and numbers of secondary outcomes.}
    \label{re_cp}
\end{figure}

Moreover, we conducted additional simulations to evaluate the efficiency gain and asymptotic inference accuracy across a range of sample sizes. Building on the two secondary outcomes settings in Case 2 (i.e., $10$ and $50$ outcomes), we added two more scenarios: 1) $5$ secondary outcomes $\{\tildeY_{1}, \tildeY_{7}, \tildeY_{11}, \tildeY_{12}, \tildeY_{13}\}$, and 2) $30$ secondary outcomes $\{\tildeY_{1}, \tildeY_{2}, \tildeY_{6}, \tildeY_{7}, \dots, \tildeY_{33}\}$. For each scenario, we examined a range of sample sizes to evaluate the empirical efficiency gain and CP of $\hatbmbeta_{PEPSI}$ and $\hatbmbeta_{avg}$. For simplicity and clarity, we only reported the results for the $\beta_2$ estimators, as similar patterns were observed for other coefficients. As shown in Figure \ref{re_cp}, $\hatbmbeta_{PEPSI}$ outperformed $\hatbmbeta_{naive}$ in efficiency once the sample size exceeded a certain threshold (e.g., $75$), and its coverage probability approached the nominal level as the sample size increased. However, integrating more secondary outcomes required a larger sample size to achieve efficiency gain and reliable coverage, due to the increased DoF. Similar trends were observed for $\hatbmbeta_{avg}$; however, it achieved more stable performance with smaller sample sizes and maintained coverage closer to the nominal level, suggesting greater robustness in small/moderate sample size settings (e.g.,  $n=100$ to $300$). Further evaluations are provided in Section 3 of the Supplementary Materials.

% \red{As verified numerically in the simulation, the sample size needs to be sufficiently larger than the total DoF of the model to facilitate appropriate inference. Recall that $K$ is the extra DoF in secondary models, defined in Theorem \ref{asymptotic pepsi}, and $p_0$ is the number of parameters in the primary model. Based on the simulation, we suggest that $n\ge 40p_0+40K$ and $n\ge 40p_0+20K$ for $\hatbmbeta_{PEPSI}$ and $\hatbmbeta_{avg}$, respectively. More evaluation and discussion in small sample size context can be found in Section ? of the Supplementary Material. 
% }

\section{Application}\label{Application}
In this case study, we investigated the effect of cigarette smoking on liver health and explored how it differed {across} age groups using the UKB data. Specifically, we considered liver PDFF and LBRS as two primary outcomes, with past smoking habits as the primary exposure. Participants were categorized into ``heavy smokers" (those who smoked most or all days) and ``others". 
% Recent literature has revealed the potential impact of smoking on chronic liver diseases \citep{marti2022cigarette}. However, the effect of smoking on fatty liver remains unclear.\red{some of these info are replicate} 
It was also of significant interest to study whether the effect of smoking differed between older and younger adults, given evidence that older smokers might have an increased risk of becoming frail \citep{kojima2018does}. Moreover, we aimed to evaluate: (1) the validity of the predicted LBRS by comparing its performance with analysis using liver PDFF as the primary outcome, and (2) the utility of PEPSI for integrating information from secondary outcomes defined from both Type 1 (PDFF) and Type 3 (LBRS) sources. 

\begin{figure}[ht]
    \centering
    \includegraphics[scale=0.3]{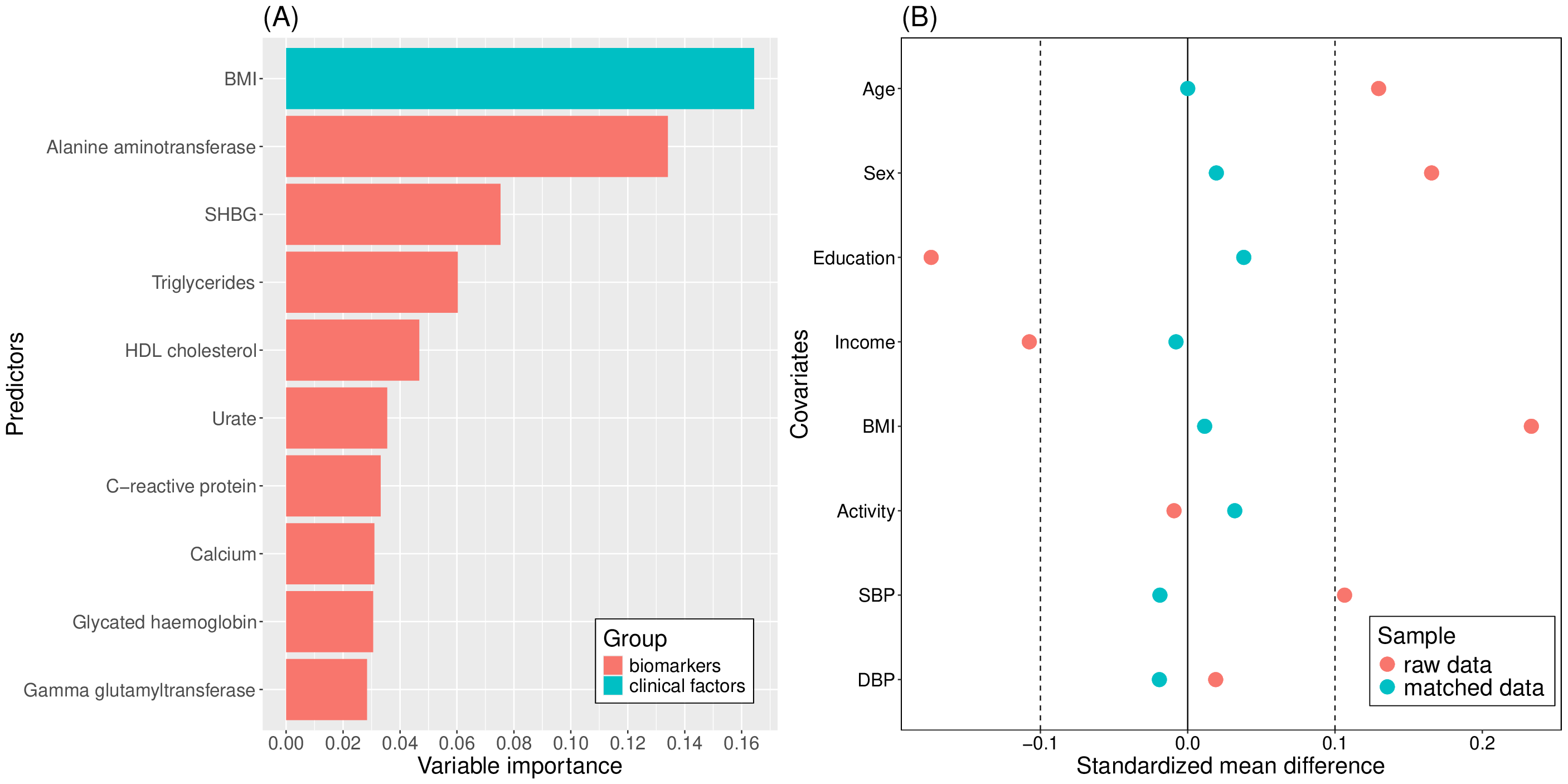}
    \caption{(A) the variable importance plot with the top 10 predictors contributing most to liver PDFF prediction. (B) the love plot assessing covariate balances of the raw and matched data.}
    \label{matched_profile}
\end{figure}

\textbf{The construction of LBRS}. To derive LBRS, we predicted liver PDFF values by building a machine learning model using clinical factors, and the blood and urine biomarkers listed in the Supplementary Materials, as predictors. Other factors were used, including a variety of baseline characteristics, such as smoking status (heavy smokers vs. others), age ($\geq$ $65$ vs. others), sex (male vs. female), education level (college or university degree vs. others), income level ($\geq$ $£31,000$ annually vs. others), BMI, sum of days performing walking, systolic blood pressure, and diastolic blood pressure, which comprehensively captured participants' demographics, lifestyle, and health conditions. 
% \textcolor{red}{Shuo Says: should we include a demographics table in the Supp for these covariates for smokers vs. non-smokers} 
The training data consisted of a cohort with complete records of liver PDFF, biomarkers and other factors, with a sample size of $2,225$. We applied XGBoost and tuned hyperparameters to minimize the mean squared error via 5-fold cross-validation. The hyperparameters included learning rates, depths of trees, the minimum number of instances in a child node, and the proportions of samples and variables in a tree. The resulting prediction model explained $26\%$ of the variability in liver PDFF variable in out of bag samples. The top $10$ important biomarkers for liver PDFF prediction are presented in Figure \ref{matched_profile}A, where BMI and alanine aminotransferase (ALT) emerged as the two most influential factors. Recent findings from another study showed that long-term high-normal ALT levels increased the risk of developing metabolic dysfunction-associated fatty liver disease (MAFLD) \citep{chen2024cumulative}, supporting the external validity of our prediction model.

\textbf{Facilitating casual interpretation}. 
% \red{I do not quite understand why you perform the matching and control the confounding variables here. We can add those confounders into primary outcome modeling, right? In the simulation, we just consider the same set of covariates for both primary and secondary outcomes. It is a little bit strange if we do this procedure here. Maybe we can provide justification why this is helpful.}
We excluded the training cohort from the downstream analysis. To unbiasedly assess the smoking effect while accounting for potential confounding factors, we applied propensity-score matching \citep{rosenbaum1983central} to align the basic profiles between ``heavy smokers" and ``others". Although matching is not required to implement our method, however, it facilitates the estimation of exposure effects within the region of common support, thereby mitigating extrapolation to areas where exposed and unexposed groups differ substantially and thus leading more unbiased comparison. To study the modifying effect of age on smoking, as suggested in the literature \citep{jang2023association}, we conducted exact matching on age group and applied propensity score matching on the remaining covariates, with propensity scores estimated using logistic regression. After matching, we obtained $1,893$ subjects in each group of ``heavy smokers" and ``others", with $239$ ($12.6\%$) older adults in each group. This matching process yielded more balanced baseline covariate distributions compared to the original, unmatched cohort (Figure \ref{matched_profile}B). The profile of the matched cohort for the final analysis are summarized in Table S8 in the Supplementary Materials.

After matching, we regressed LBRS on smoking status, age, and their interaction, and then calculated the marginal effect of smoking on LBRS, as well as the subgroup effects for different age groups. We applied the PEPSI framework for parameter estimation by integrating information from the secondary outcomes (i.e., blood and urine biomarkers). OLS regression was used for both the primary and secondary data analysis, and we compared the PEPSI estimator with a naive OLS estimator that did not integrate information from secondary data as a benchmark. For both estimators, we calculated 95\% confidence intervals and p-values from Wald tests against the null hypothesis that the underlying values were zeros. In addition, we considered E-value to explore the potential impact of unmeasured confounders \citep{vanderweele2017sensitivity}, such as dietary preference. 

\begin{table}[ht]
\centering
\caption{Results of the naive OLS and PEPSI methods for estimating the smoking effect on LBRS.
% \red{coding for smoking and age (categorical variables?). I think you may want to have a better code on the variables.}
}
\label{main_application_result}
\resizebox{\textwidth}{!}{%
\begin{tabular}{@{}ccccccccccccc@{}}
\toprule
 &  & \multicolumn{5}{c}{OLS} & \multicolumn{6}{c}{PEPSI} \\
 &  & EST & SE & LL & UL & P & EST & SE & LL & UL & P & RE \\ \midrule
\multirow{4}{*}{\begin{tabular}[c]{@{}c@{}}Regression\\ parameters\end{tabular}} & Intercept & 5.526 & 0.077 & 5.376 & 5.677 & 0.000 & 5.499 & 0.057 & 5.387 & 5.611 & 0.000 & 1.813 \\
 & Smoking & 0.055 & 0.106 & -0.154 & 0.264 & 0.605 & 0.063 & 0.052 & -0.040 & 0.166 & 0.229 & 4.138 \\
 & Smoking:Age & 0.084 & 0.267 & -0.438 & 0.607 & 0.752 & 0.281 & 0.118 & 0.049 & 0.513 & 0.018 & 5.087 \\
 & Age & -0.211 & 0.187 & -0.579 & 0.156 & 0.259 & -0.227 & 0.093 & -0.410 & -0.044 & 0.015 & 4.030 \\
 &  &  &  &  &  &  &  &  &  &  &  &  \\
\multirow{3}{*}{\begin{tabular}[c]{@{}c@{}}Effect of\\ smoking\end{tabular}} & Age below 65 & 0.055 & 0.106 & -0.154 & 0.264 & 0.605 & 0.063 & 0.052 & -0.040 & 0.166 & 0.229 & 4.138 \\
 & Age no less than 65 & 0.139 & 0.245 & -0.340 & 0.619 & 0.569 & 0.344 & 0.108 & 0.132 & 0.555 & 0.001 & 5.137 \\
 & Overall & 0.066 & 0.098 & -0.126 & 0.258 & 0.503 & 0.098 & 0.048 & 0.004 & 0.193 & 0.041 & 4.140 \\ \bottomrule
\end{tabular}%
}
\begin{tablenotes}[flushleft]
\item \footnotesize Abbreviations: EST, parameter estimates; SE, estimated standard errors; LL \& UL, $95\%$ lower and upper confidence limits of estimates; P, p-values; RE, relative efficiency, ratios of estimated variances between naive OLS and PEPSI model estimates.
\end{tablenotes}
\end{table}

\textbf{Results interpretation}.  The results are summarized in Table \ref{main_application_result}. Compared with the naive OLS estimator, the PEPSI estimator yielded similar point estimates, implying that the proposed integrative learning does not introduce estimation bias to the original analysis. However, the PEPSI estimator showed substantial variance reduction and narrower confidence intervals, with RE greater than $4$ for many estimates. This signifies the utility of the PEPSI estimator when integrating information from secondary outcomes. Notably, the PEPSI estimator identified more statistically significant factors, whereas the naive OLS estimator failed to detect any significant association. In particular, the PEPSI results showed that the LBRS was significantly lower in older adults compared to younger adults when controlling for the same smoking habit. Regarding cigarette smoking, we observed significantly higher LBRS values in heavy smokers. A more detailed analysis revealed that cigarette smoking had a positive and significant effect on LBRS in older adults, and this effect was significantly larger than that in younger adults. In contrast, cigarette smoking might not have a significant impact on LBRS among younger adults. These findings suggest that cigarette smoking affects LBRS, with hetreogeneous effects across different sub-populations. Furthermore, the E-values for the smoking effects were $1.45$ for adults aged $65$ or older and $1.21$ for the general population. These values imply that our analysis exhibit moderate/mild robustness against unmeasured confounders \citep{vanderweele2017sensitivity}.

% As shown in Table \ref{main_application_result}, the PEPSI model has much smaller SE and narrower confidence intervals, resulting in identifying more statistically significant factors, while the naive OLS model fails to identify any. 

\begin{table}[ht]
\centering
\caption{Analysis results of the naive OLS and PEPSI methods for estimating the smoking effect on raw liver PDFF.}
\label{application_result_sensitivity}
\resizebox{\textwidth}{!}{%
\begin{tabular}{@{}ccccccccccccc@{}}
\toprule
 &  & \multicolumn{5}{c}{OLS} & \multicolumn{6}{c}{PEPSI} \\
 &  & EST & SE & LL & UL & P & EST & SE & LL & UL & P & RE \\ \midrule
\multirow{4}{*}{\begin{tabular}[c]{@{}c@{}}Regression\\ parameters\end{tabular}} & Intercept & 5.706 & 0.301 & 5.117 & 6.295 & 0.000 & 5.642 & 0.285 & 5.083 & 6.201 & 0.000 & 1.111 \\
 & Smoking & -0.195 & 0.413 & -1.003 & 0.614 & 0.637 & -0.157 & 0.376 & -0.894 & 0.581 & 0.678 & 1.202 \\
 & Smoking:Age & 2.537 & 0.977 & 0.621 & 4.452 & 0.009 & 1.829 & 0.800 & 0.261 & 3.398 & 0.022 & 1.491 \\
 & Age & -1.071 & 0.534 & -2.118 & -0.024 & 0.045 & -0.785 & 0.470 & -1.706 & 0.137 & 0.095 & 1.291 \\
 &  &  &  &  &  &  &  &  &  &  &  &  \\
\multirow{3}{*}{\begin{tabular}[c]{@{}c@{}}Effect of\\ smoking\end{tabular}} & Age below 65 & -0.195 & 0.413 & -1.003 & 0.614 & 0.637 & -0.157 & 0.376 & -0.894 & 0.581 & 0.678 & 1.202 \\
 & Age no less than 65 & 2.342 & 0.886 & 0.606 & 4.078 & 0.008 & 1.673 & 0.709 & 0.284 & 3.062 & 0.018 & 1.561 \\
 & Overall & 0.187 & 0.375 & -0.549 & 0.922 & 0.619 & 0.118 & 0.338 & -0.544 & 0.781 & 0.726 & 1.232 \\ \bottomrule
\end{tabular}%
}
\begin{tablenotes}[flushleft]
\item \footnotesize Abbreviations: EST, parameter estimates; SE, estimated standard errors; LL \& UL, $95\%$ lower and upper confidence limits of estimates; P, p-values; RE, relative efficiency, ratios of estimated variances between naive OLS and PEPSI model estimates.
\end{tablenotes}
\end{table}

\textbf{Results validation}. We validated the findings by repeating the analysis using liver PDFF as the primary outcome, while keeping all other procedures identical to those previously described. A total of $1,078$ subjects were included in this analysis (demographics were similar to the primary analysis, see Table S9 in the Supplementary Materials), and the results are summarized in Table \ref{application_result_sensitivity}. Most estimates showed consistent patterns with Table \ref{main_application_result}, such as significant smoking effects in older adults but not in younger individuals. These findings validate our findings above and suggest that LBRS can serve as a reliable surrogate outcome for the raw liver PDFF data. However, the estimated effect sizes were larger when using liver PDFF. Regarding efficiency gain, the RE was less pronounced than in the analysis using LBRS, which is expected given the weaker associations between the primary outcome (liver PDFF) and the secondary outcomes. Nevertheless, the PEPSI estimates were still more efficient than the OLS estimates, with some RE values reaching as high as $1.6$. Additionally, we conducted two supplementary analysis: one in which liver PDFF values were predicted only by the blood and urine biomarkers, and another external validation analysis using the Alzheimer's Disease Neuroimaging Initiative dataset \citep{mueller2005alzheimer}. Both analysis yielded conclusions consistent with our primary findings and are presented in Tables S10 and S14 in the Supplementary Materials.

\section{Discussion}\label{Discussion}

In this paper, we have developed a new integrative learning framework that enhances inferences on the primary outcome by leveraging additional information from multiple cross-sectional secondary outcomes. The simulation/example code is available online (\url{github.com/daxuand1/pepsi}). 
% This method is not only computationally efficient and optimal within a class of estimators but also robust against misspecifications in secondary models and superior to existing integration schemes, as evidenced by our simulation studies. 
In our case study using UKB data, this approach significantly improved the analysis, revealing that cigarette smoking may adversely affect liver health, with particularly stronger effects observed in older adults. Notably, our findings suggest that older adult smokers may have a higher risk of developing fatty liver, while younger smokers may not experience the same impact. This discrepancy could be due to older adults having longer smoking histories or multiple pathways, including toxic, immunologic, and oncogenic factors, making them more susceptible as their biological systems become less resilient with age \citep{kojima2018does}. 
% \textcolor{red}{Shuo Says: Smoking may increase the risk of fatty liver through multiple pathways, for example, . These pathways of older adults can be more sensitive to smoking as the biological systems become less resilient during aging.}
These results enrich the current knowledge about the detrimental effects of cigarette smoking on liver health \citep{marti2022cigarette} and support public health recommendations for older adults to quit smoking. Furthermore, by comparing the results obtained from the predicted LBRS with those from the original liver PDFF measure, we observed reduced covariate effects, indicating that the predicted risk score may lose some information from the original outcome, leading to decreased statistical power for detecting covariate significance. Thus, integrating information from individual predictors as secondary outcomes is a natural and promising approach to enhance the statistical performance of risk prediction score analysis. Additionally, the efficiency gain was substantially larger for the PEPSI estimator when the predicted LBRS serves as the primary outcome compared to the original liver PDFF outcome. This finding implies that, in practice, the PEPSI estimator is especially favorable and strongly recommended when secondary outcomes are obtained from Type 2 or 3 sources.

Extensive simulation studies suggest that the proposed methods perform well when $n\ge 40p_0+40K$ and $n\ge 40p_0+20K$ for $\hatbmbeta_{PEPSI}$ and $\hatbmbeta_{avg}$, respectively, where $p_0$ is the number of parameters in the primary model, and $K$ is the extra DoF in the secondary models (defined in Theorem \ref{asymptotic pepsi}). These observations provide practical guidance for study design. Moreover, our simulations suggest that for sample sizes around $300$ or fewer, no more than $10$ secondary
outcomes should be used with $\hatbmbeta_{PEPSI}$; for sample sizes between $300$ and $500$, up to  $30$ secondary outcomes can be integrated; and for sample sizes of $500$ or above, $\hatbmbeta_{PEPSI}$ can accommodate $50$ secondary outcomes. Otherwise, $\hatbmbeta_{avg}$ is preferred over $\hatbmbeta_{PEPSI}$. Further evaluations are needed to develop more comprehensive guidelines for practical use.

From a methodological standpoint, there are several potential extensions of the PEPSI framework. First, while this paper focuses on low-dimensional covariates as an initial study of PEPSI, it would be valuable to explore extensions to high-dimensional settings. Second, the current framework is based on the generalized linear model, which could be adapted for use with missing data, time-to-event models, or causal inference. Third, beyond integrating data within a single study, adapting the PEPSI framework for information integration across multiple independent datasets holds great promise.  Fourth, it would be interesting to develop an empirical likelihood ratio test within the proposed framework. In theory, it may be possible to develop an ELR test by stacking the estimating equations from both the primary and secondary outcomes and jointly estimating the parameters under a penalized empirical likelihood framework \citep{leng2012penalized}. However, as discussed in Section \ref{Multiple Secondary Data}, this approach may suffer from high-dimensional estimating equations and can be computationally intensive and unstable. These extensions require considerable effort and warrant further research.

Despite its methodological advantages, we acknowledge two theoretical challenges that remain unaddressed in this paper. First, although we have provided a useful guideline for selecting secondary outcomes (Section \ref{Multiple Secondary Data}), quantifying uncertainty associated with this selection process is necessary but remains challenging. Nonetheless, it is important to emphasize that the selection of secondary outcomes does not alter the target population and the underlying effect size defined by the primary outcome and original study design. Second, when applying the PEPSI estimator with a very large number of secondary outcomes (e.g., thousands), the standard errors may be under-estimated. This issue can be partially attributed to uncertainties arising from implementing PCA and tuning parameter selection in numerous penalized optimizations. Addressing these challenges requires more advanced theoretical development, which is beyond the scope of this paper.

\section{Competing interests}
No competing interest is declared.

\section{Acknowledgments}
The authors are grateful for the detailed comments of referees and editors on the previous version of the paper.

\section{Supplementary Material}
Supplementary material is available at Journal of the Royal Statistical Society: Series B online.

\section{Funding}
C. Chen's work was partly supported by the National Institutes of Health (NIH) grants P30AG028747, R01AG089377 and Johns Hopkins Institute for Clinical and Translational Research (JHU-ICTR) funded by 1UM1TR004926. M. Wang's work was partly supported by NIH Grant R01HL175410.  The contents are solely the responsibility of the authors and do not necessarily represent the official view of NIH or JHU-ICTR.

\section{Data availability}
The UK Biobank data are publicly available to approved researchers through the UK Biobank Access Management System (https://www.ukbiobank.ac.uk). The ADNI data are publicly accessible through the Alzheimer’s Disease Neuroimaging Initiative data repository (https://adni.loni.usc.edu), subject to standard data-use agreements.

\bibliographystyle{abbrvnat}
\bibliography{bibliography}

\begin{thebibliography}{38}
\providecommand{\natexlab}[1]{#1}
\providecommand{\url}[1]{\texttt{#1}}
\expandafter\ifx\csname urlstyle\endcsname\relax
  \providecommand{\doi}[1]{doi: #1}\else
  \providecommand{\doi}{doi: \begingroup \urlstyle{rm}\Url}\fi

\bibitem[Alsefri et~al.(2020)Alsefri, Sudell, Garc{\'\i}a-Fi{\~n}ana, and
  Kolamunnage-Dona]{alsefri2020bayesian}
M.~Alsefri, M.~Sudell, M.~Garc{\'\i}a-Fi{\~n}ana, and R.~Kolamunnage-Dona.
\newblock Bayesian joint modelling of longitudinal and time to event data: a
  methodological review.
\newblock \emph{BMC medical research methodology}, 20:\penalty0 1--17, 2020.

\bibitem[Bandeen-Roche et~al.(2015)Bandeen-Roche, Seplaki, Huang, Buta,
  Kalyani, Varadhan, Xue, Walston, and Kasper]{bandeen2015frailty}
K.~Bandeen-Roche, C.~L. Seplaki, J.~Huang, B.~Buta, R.~R. Kalyani, R.~Varadhan,
  Q.-L. Xue, J.~D. Walston, and J.~D. Kasper.
\newblock Frailty in older adults: a nationally representative profile in the
  united states.
\newblock \emph{Journals of Gerontology Series A: Biomedical Sciences and
  Medical Sciences}, 70\penalty0 (11):\penalty0 1427--1434, 2015.

\bibitem[Breitling(2015)]{breitling2015smoking}
L.~P. Breitling.
\newblock Smoking as an effect modifier of the association of calcium intake
  with bone mineral density.
\newblock \emph{The Journal of Clinical Endocrinology \& Metabolism},
  100\penalty0 (2):\penalty0 626--635, 2015.

\bibitem[Chatterjee et~al.(2016)Chatterjee, Chen, Maas, and
  Carroll]{chatterjee2016constrained}
N.~Chatterjee, Y.-H. Chen, P.~Maas, and R.~J. Carroll.
\newblock Constrained maximum likelihood estimation for model calibration using
  summary-level information from external big data sources.
\newblock \emph{Journal of the American Statistical Association}, 111\penalty0
  (513):\penalty0 107--117, 2016.

\bibitem[Chen et~al.(2022)Chen, Han, and He]{chen2022improving}
C.~Chen, P.~Han, and F.~He.
\newblock Improving main analysis by borrowing information from auxiliary data.
\newblock \emph{Statistics in Medicine}, 41\penalty0 (3):\penalty0 567--579,
  2022.

\bibitem[Chen et~al.(2023{\natexlab{a}})Chen, Wang, and
  Chen]{chen2023efficient}
C.~Chen, M.~Wang, and S.~Chen.
\newblock An efficient data integration scheme for synthesizing information
  from multiple secondary datasets for the parameter inference of the main
  analysis.
\newblock \emph{Biometrics}, 2023{\natexlab{a}}.

\bibitem[Chen et~al.(2023{\natexlab{b}})Chen, Yu, Shen, and
  Wang]{chen2023synthesizing}
C.~Chen, T.~Yu, B.~Shen, and M.~Wang.
\newblock Synthesizing secondary data into survival analysis to improve
  estimation efficiency.
\newblock \emph{Biometrical Journal}, 65\penalty0 (3):\penalty0 2100326,
  2023{\natexlab{b}}.

\bibitem[Chen et~al.(2024{\natexlab{a}})Chen, Han, Chen, Shardell, and
  Qin]{chen2024integrating}
C.~Chen, P.~Han, S.~Chen, M.~Shardell, and J.~Qin.
\newblock Integrating external summary information in the presence of prior
  probability shift: an application to assessing essential hypertension.
\newblock \emph{Biometrics}, 80\penalty0 (3):\penalty0 ujae090,
  2024{\natexlab{a}}.

\bibitem[Chen et~al.(2024{\natexlab{b}})Chen, Wu, Liu, Yan, Wang, Xing, Song,
  and Ding]{chen2024cumulative}
J.-F. Chen, Z.-Q. Wu, H.-S. Liu, S.~Yan, Y.-X. Wang, M.~Xing, X.-Q. Song, and
  S.-Y. Ding.
\newblock Cumulative effects of excess high-normal alanine aminotransferase
  levels in relation to new-onset metabolic dysfunction-associated fatty liver
  disease in china.
\newblock \emph{World Journal of Gastroenterology}, 30\penalty0 (10):\penalty0
  1346, 2024{\natexlab{b}}.

\bibitem[Deng et~al.(2024)Deng, Chinchilli, Feng, Chen, and
  Wang]{deng2024robust}
D.~Deng, V.~M. Chinchilli, H.~Feng, C.~Chen, and M.~Wang.
\newblock Robust integration of secondary outcomes information into primary
  outcome analysis in the presence of missing data.
\newblock \emph{Statistical Methods in Medical Research}, page
  09622802241254195, 2024.

\bibitem[Duan et~al.(2022)Duan, Ning, and Chen]{duan2022heterogeneity}
R.~Duan, Y.~Ning, and Y.~Chen.
\newblock Heterogeneity-aware and communication-efficient distributed
  statistical inference.
\newblock \emph{Biometrika}, 109\penalty0 (1):\penalty0 67--83, 2022.

\bibitem[Egan et~al.(2016)Egan, Li, and Wagner]{egan2016systolic}
B.~M. Egan, J.~Li, and C.~S. Wagner.
\newblock Systolic blood pressure intervention trial (sprint) and target
  systolic blood pressure in future hypertension guidelines.
\newblock \emph{Hypertension}, 68\penalty0 (2):\penalty0 318--323, 2016.

\bibitem[Ezzalfani et~al.(2019)Ezzalfani, Burzykowski, and
  Paoletti]{ezzalfani2019joint}
M.~Ezzalfani, T.~Burzykowski, and X.~Paoletti.
\newblock Joint modelling of a binary and a continuous outcome measured at two
  cycles to determine the optimal dose.
\newblock \emph{Journal of the Royal Statistical Society Series C: Applied
  Statistics}, 68\penalty0 (2):\penalty0 369--384, 2019.

\bibitem[Fan and Li(2001)]{fan2001variable}
J.~Fan and R.~Li.
\newblock Variable selection via nonconcave penalized likelihood and its oracle
  properties.
\newblock \emph{Journal of the American statistical Association}, 96\penalty0
  (456):\penalty0 1348--1360, 2001.

\bibitem[Han et~al.(2024)Han, Li, Park, Mukherjee, and
  Taylor]{han2024improving}
P.~Han, H.~Li, S.~K. Park, B.~Mukherjee, and J.~M. Taylor.
\newblock Improving prediction of linear regression models by integrating
  external information from heterogeneous populations: James--stein estimators.
\newblock \emph{Biometrics}, 80\penalty0 (3), 2024.

\bibitem[Jang et~al.(2023)Jang, Joo, Park, Park, and Jang]{jang2023association}
Y.~S. Jang, H.~J. Joo, Y.~S. Park, E.-C. Park, and S.-I. Jang.
\newblock Association between smoking cessation and non-alcoholic fatty liver
  disease using nafld liver fat score.
\newblock \emph{Frontiers in Public Health}, 11:\penalty0 1015919, 2023.

\bibitem[Ke et~al.(2017)Ke, Meng, Finley, Wang, Chen, Ma, Ye, and
  Liu]{ke2017lightgbm}
G.~Ke, Q.~Meng, T.~Finley, T.~Wang, W.~Chen, W.~Ma, Q.~Ye, and T.-Y. Liu.
\newblock Lightgbm: A highly efficient gradient boosting decision tree.
\newblock \emph{Advances in neural information processing systems}, 30, 2017.

\bibitem[Kojima et~al.(2018)Kojima, Iliffe, Jivraj, Liljas, and
  Walters]{kojima2018does}
G.~Kojima, S.~Iliffe, S.~Jivraj, A.~Liljas, and K.~Walters.
\newblock Does current smoking predict future frailty? the english longitudinal
  study of ageing.
\newblock \emph{Age and ageing}, 47\penalty0 (1):\penalty0 126--131, 2018.

\bibitem[Kramer et~al.(2017)Kramer, Pickhardt, Kliewer, Hernando, Chen,
  Zagzebski, and Reeder]{kramer2017accuracy}
H.~Kramer, P.~J. Pickhardt, M.~A. Kliewer, D.~Hernando, G.-H. Chen, J.~A.
  Zagzebski, and S.~B. Reeder.
\newblock Accuracy of liver fat quantification with advanced ct, mri, and
  ultrasound techniques: prospective comparison with mr spectroscopy.
\newblock \emph{American journal of Roentgenology}, 208\penalty0 (1):\penalty0
  92--100, 2017.

\bibitem[Kwo et~al.(2017)Kwo, Cohen, and Lim]{kwo2017acg}
P.~Y. Kwo, S.~M. Cohen, and J.~K. Lim.
\newblock Acg clinical guideline: evaluation of abnormal liver chemistries.
\newblock \emph{Official journal of the American College of Gastroenterology|
  ACG}, 112\penalty0 (1):\penalty0 18--35, 2017.

\bibitem[Leng and Tang(2012)]{leng2012penalized}
C.~Leng and C.~Y. Tang.
\newblock Penalized empirical likelihood and growing dimensional general
  estimating equations.
\newblock \emph{Biometrika}, 99\penalty0 (3):\penalty0 703--716, 2012.

\bibitem[Levit et~al.(2024)Levit, Garrett-Mayer, Peppercorn, and
  Ratain]{levit2024critical}
L.~A. Levit, E.~Garrett-Mayer, J.~Peppercorn, and M.~J. Ratain.
\newblock Critical importance of correctly defining and reporting secondary
  endpoints when assessing the ethics of research biopsies.
\newblock \emph{Clinical Trials}, page 17407745241244753, 2024.

\bibitem[Longo et~al.(2012)]{longo2012harrisons}
D.~L. Longo et~al.
\newblock \emph{Harrison's principles of internal medicine}.
\newblock Biblioteca Hern{\'a}n Malo Gonz{\'a}lez, 2012.

\bibitem[Marti-Aguado et~al.(2022)Marti-Aguado, Clemente-Sanchez, and
  Bataller]{marti2022cigarette}
D.~Marti-Aguado, A.~Clemente-Sanchez, and R.~Bataller.
\newblock Cigarette smoking and liver diseases.
\newblock \emph{Journal of hepatology}, 77\penalty0 (1):\penalty0 191--205,
  2022.

\bibitem[Mueller et~al.(2005)Mueller, Weiner, Thal, Petersen, Jack, Jagust,
  Trojanowski, Toga, and Beckett]{mueller2005alzheimer}
S.~G. Mueller, M.~W. Weiner, L.~J. Thal, R.~C. Petersen, C.~Jack, W.~Jagust,
  J.~Q. Trojanowski, A.~W. Toga, and L.~Beckett.
\newblock The alzheimer's disease neuroimaging initiative.
\newblock \emph{Neuroimaging Clinics}, 15\penalty0 (4):\penalty0 869--877,
  2005.

\bibitem[Owen(2001)]{owen2001empirical}
A.~B. Owen.
\newblock \emph{Empirical likelihood}.
\newblock CRC press, 2001.

\bibitem[Qin and Lawless(1994)]{qin1994empirical}
J.~Qin and J.~Lawless.
\newblock Empirical likelihood and general estimating equations.
\newblock \emph{the Annals of Statistics}, 22\penalty0 (1):\penalty0 300--325,
  1994.

\bibitem[Rodrigue et~al.(2023)Rodrigue, Hayes, Waite, Corcoran, Glahn, and
  Jalbrzikowski]{rodrigue2023multimodal}
A.~L. Rodrigue, R.~A. Hayes, E.~Waite, M.~Corcoran, D.~C. Glahn, and
  M.~Jalbrzikowski.
\newblock Multimodal neuroimaging summary scores as neurobiological markers of
  psychosis.
\newblock \emph{Schizophrenia Bulletin}, page sbad149, 2023.

\bibitem[Rosenbaum and Rubin(1983)]{rosenbaum1983central}
P.~R. Rosenbaum and D.~B. Rubin.
\newblock The central role of the propensity score in observational studies for
  causal effects.
\newblock \emph{Biometrika}, 70\penalty0 (1):\penalty0 41--55, 1983.

\bibitem[Starekova et~al.(2021)Starekova, Hernando, Pickhardt, and
  Reeder]{starekova2021quantification}
J.~Starekova, D.~Hernando, P.~J. Pickhardt, and S.~B. Reeder.
\newblock Quantification of liver fat content with ct and mri: state of the
  art.
\newblock \emph{Radiology}, 301\penalty0 (2):\penalty0 250--262, 2021.

\bibitem[Van~der Vaart(2000)]{van2000asymptotic}
A.~W. Van~der Vaart.
\newblock \emph{Asymptotic statistics}, volume~3.
\newblock Cambridge university press, 2000.

\bibitem[VanderWeele and Ding(2017)]{vanderweele2017sensitivity}
T.~J. VanderWeele and P.~Ding.
\newblock Sensitivity analysis in observational research: introducing the
  e-value.
\newblock \emph{Annals of internal medicine}, 167\penalty0 (4):\penalty0
  268--274, 2017.

\bibitem[Wolf et~al.(2024)Wolf, Koopmeiners, and Vock]{wolf2024commentary}
J.~M. Wolf, J.~S. Koopmeiners, and D.~M. Vock.
\newblock Commentary on chen et al.(2022): The need for continued
  methodological research on leveraging information in secondary endpoints for
  more efficient rcts.
\newblock \emph{Contemporary Clinical Trials}, 145:\penalty0 107664, 2024.

\bibitem[Xia et~al.(2023)Xia, Du, Li, Wang, Zha, Wu, and
  Ju]{xia2023association}
T.~Xia, M.~Du, H.~Li, Y.~Wang, J.~Zha, T.~Wu, and S.~Ju.
\newblock Association between liver mri proton density fat fraction and liver
  disease risk.
\newblock \emph{Radiology}, 309\penalty0 (1):\penalty0 e231007, 2023.

\bibitem[Zhai and Han(2024)]{zhai2024integrating}
Y.~Zhai and P.~Han.
\newblock Integrating external summary information under population
  heterogeneity and information uncertainty.
\newblock \emph{Electronic Journal of Statistics}, 18\penalty0 (2):\penalty0
  5304--5329, 2024.

\bibitem[Zhang(2010)]{zhang2010nearly}
C.-H. Zhang.
\newblock Nearly unbiased variable selection under minimax concave penalty.
\newblock \emph{The Annals of Statistics}, pages 894--942, 2010.

\bibitem[Zhang et~al.(2020)Zhang, Deng, Schiffman, Qin, and
  Yu]{zhang2020generalized}
H.~Zhang, L.~Deng, M.~Schiffman, J.~Qin, and K.~Yu.
\newblock Generalized integration model for improved statistical inference by
  leveraging external summary data.
\newblock \emph{Biometrika}, 107\penalty0 (3):\penalty0 689--703, 2020.

\bibitem[Zou(2006)]{zou2006adaptive}
H.~Zou.
\newblock The adaptive lasso and its oracle properties.
\newblock \emph{Journal of the American statistical association}, 101\penalty0
  (476):\penalty0 1418--1429, 2006.

\end{thebibliography}

\end{document}